\PassOptionsToPackage{pdfpagelabels=false}{hyperref} 
\documentclass[fleqn,usenatbib]{mnras}

\usepackage{newtxtext,newtxmath}

\usepackage[utf8]{inputenc} 
\usepackage[T1]{fontenc}    
\usepackage{hyperref}       
\usepackage{url}            
\usepackage[table]{xcolor}

\usepackage{graphicx}	
\usepackage{amsmath}	
\usepackage{amssymb}	
\usepackage{bm}

\usepackage{breakcites}
\usepackage{url} 
\usepackage{float}
\usepackage{subfigure}
\usepackage{color}
\usepackage{lineno} 
\usepackage{enumitem}  
\usepackage{chngcntr}
\usepackage{cancel}
\usepackage{threeparttable}
\usepackage{tikz}
\usepackage{academicons}

\usepackage{pifont}
\newcommand{\xmark}{\ding{55}}%

\newcommand{\eqb}{\begin{eqnarray}}
\newcommand{\eqe}{\end{eqnarray}}
\newcommand{\bi}{\begin{itemize}}
\newcommand{\ei}{\end{itemize}}

\definecolor{cornflowerblue}{rgb}{0.39, 0.58, 0.93}
\definecolor{darkspringgreen}{rgb}{0.09, 0.45, 0.27}
\definecolor{turquoise}{rgb}{0.03, 0.91, 0.87}
\definecolor{coral}{rgb}{0.97, 0.51, 0.47}
\definecolor{indian}{rgb}{1.0, 0.22, 0.0}
\definecolor{darkorchid}{rgb}{0.6, 0.2, 0.8}
\definecolor{lime}{rgb}{0.8, 1.0, 0.0}

\definecolor{lime}{HTML}{A6CE39}
\DeclareRobustCommand{\orcidicon}{%
	\begin{tikzpicture}
	\draw[lime, fill=lime] (0,0) 
	circle [radius=0.16] 
	node[white] {{\fontfamily{qag}\selectfont \tiny ID}};
	\draw[white, fill=white] (-0.0625,0.095) 
	circle [radius=0.007];
	\end{tikzpicture}
	\hspace{-2mm}
}

\foreach \x in {A, ..., Z}{%
	\expandafter\xdef\csname orcid\x\endcsname{\noexpand\href{https://orcid.org/\csname orcidauthor\x\endcsname}{\noexpand\orcidicon}}
}

\newcommand{\orcid}[1]{\href{https://orcid.org/#1}{\textcolor[HTML]{A6CE39}{\orcidicon}}}
\hypersetup{colorlinks,linkcolor={teal},citecolor={teal},urlcolor={teal}}  

\title[HSC in expanding sources]{Hadronic supercriticality in spherically expanding sources: application to GRB prompt emission}
\author[Florou, Mastichiadis, Petropoulou]{
Ioulia Florou\orcid{0000-0002-7708-4041}$^{1}$\thanks{E-mail: iflorou@phys.uoa.gr},
Apostolos Mastichiadis\orcid{0000-0001-5217-4801}$^{1}$\thanks{E-mail: amastich.phys.uoa.gr}
Maria Petropoulou\orcid{0000-0001-6640-0179}$^{1}$
\thanks{Mercator Fellow}
\\
$^{1}$Department of Physics, National and Kapodistrian University of Athens, University Campus Zografos, GR 15783, Greece\\
}

\date{Accepted XXX. Received YYY; in original form ZZZ}

\begin{document}

\label{firstpage}
\pagerange{\pageref{firstpage}--\pageref{lastpage}}
\maketitle

\begin{abstract}
Relativistic hadronic plasmas  can become under certain conditions supercritical, abruptly and efficiently releasing the energy stored in protons through photon outbursts. Past studies have tried to relate the features of such hadronic supercriticalities (HSC) to the phenomenology of Gamma-Ray Burst (GRB) prompt emission. In this work we investigate, for the first time, HSC in adiabatically expanding sources. We examine the conditions required to trigger HSC, study the role of expansion velocity, and discuss our results in relation to GRB prompt emission. We find multi-pulse light curves from slowly expanding regions ($u_{\rm exp}\lesssim 0.01 c)$ that are a manifestation of the natural HSC quasi-periodicity, while single-pulse light curves with a fast rise and slow decay are found for higher velocities. The formation of the photon spectrum is governed by an in-source electromagnetic cascade. The peak photon energy is $\sim 1$~MeV ($\sim 1$~GeV) for maximum proton energies $\sim 1-10$ PeV ($1-10$~EeV) assuming a jet Lorentz factor 100. Peak $\gamma$-ray luminosities are in the range $10^{49}-10^{52}$~erg s$^{-1}$, with the MeV-peaked spectra being $\sim 100-300$ times more luminous than their GeV-peaked analogues. HSC bursts peaking in the MeV are also copious $\sim 10$ TeV neutrino emitters, with an all-flavour fluence $\sim 10\%$ of the $\gamma$-ray one. The hypothesis that typical long-duration GRBs are powered by HSC could be tested in the near future with more sensitive neutrino telescopes like IceCube-Gen2.
\end{abstract}




\section{Introduction}

Gamma Ray Bursts (GRBs) are brief flashes of $\gamma$-rays and are considered to be one of the most energetic transient explosive phenomena in the universe. The production mechanism of a GRB remains still a challenging problem in high-energy astrophysics since their first discovery more than a half century ago. The main phase of the  GRB phenomenon, namely the prompt emission phase, is associated with several notable characteristics, such as the high photon luminosity and highly variable light curves \footnote{In some cases the GRB light curves consist of a single pulse that shows a Fast Rise and an Exponential Decay
(FRED).},  usually consisting of many pulses, each of them lasting from 1 msec to 1 sec. The GRB prompt emission can last in total from seconds to minutes, i.e., less than 2~s in the case of short GRBs and more than 100~s in the case of  long GRBs, reaching luminosities up to $\rm L_{\gamma}=10^{54}$ erg~s$^{-1}$ (for reviews see  \cite{KUMAR2015,Beloborodov_2017}).  
In the last several years, a number of GRB events with considerable longer duration has been detected. These outbursts, named Ultra Long GRBs (ULGRBs), are characterised by a $\gamma$-ray emission that lasts for several thousands of seconds \citep{Gendre_2013}.
In general, the spectrum of prompt emission has a well defined peak at an energy $\varepsilon_{\rm pk}$ that  typically lies in the range of $0.1-1$ MeV.  Some years ago the Fermi Large Area Telescope (LAT) detected GeV photons from several GRBs and in some cases it found additional spectral components in the GeV energy range \citep{2009Abdo,Ackermann_2011,Goldstein_2012}. A general feature is that the onset of the GeV emission tends to be delayed up to some seconds, relatively to the onset of the main MeV emission, suggesting a likely different origin of these photons. More recently, ground-based Cherenkov telescopes have detected even higher energy photons, i.e., greater than 100 GeV,  in the early afterglow of certain GRBs \citep{2019magiccoll,Abdalla_2019}, suggesting that GRBs can be one of the most extreme astrophysical accelerators in the universe.

The majority of the GRB spectra in the prompt emission phase are well fitted by the so-called Band function \citep{Band}, which consists of two power laws smoothly connected at the energy $\varepsilon_{\rm pk}$.  Although various models have been proposed during the past few decades, the radiation mechanisms responsible for the prompt GRB emission remain still a puzzle. Generally, there are two classes of models for GRB emission depending on the species of the radiating particles, namely the leptonic and the hadronic ones. 
In the leptonic scenario,  electron synchrotron radiation  was one of the first mechanisms proposed to explain the prompt GRB emission \citep{1994katz,Meszaros_1994,1996Sari-Nar-Piran,Sari_1998}.  However, the optically thin electron synchrotron model faced several theoretical difficulties, such as the \textit{line of death} problem, that refers to the discrepancy between the observed and model predicted low energy photon index. \citep{Crider_1997,Preece98}. A refined time-dependent spectral analysis in several bright bursts \citep{2012Guiriec,Oganesyan17,Oganesyan19} has shown that the GRB spectra may be better fitted with a multi-component model or with a phenomenological function consisting of two broken power laws instead of the classical Band function. The availability of more detailed spectral information has led the GRB community to readdress the electron synchrotron model as the main radiative mechanism responsible for the prompt emission for the majority of GRBs. \cite{2020burges} have performed a time resolved spectral analysis to the prompt spectra of single pulse GRBs, detected by the FERMI Gamma Ray Monitor (GBM),  and have concluded that the electron synchrotron interpretation is a feasible option once time dependence and cooling are properly included.  

Proton synchrotron radiation was also proposed to explain the prompt GRB emission \citep{ghisselini20}, motivated by recent spectral analysis of GRBs detected by the Swift Burst Alert Telescope (BAT) \citep{Oganesyan17,Oganesyan19}. However, it was recently demonstrated that the proposed scenario cannot explain the observed GRB spectra, unless very high bulk Lorentz factors ($>1000$) are assumed, implying a very high required jet power, i.e. $\gtrsim 10^{54}$ erg s$^{-1}$ \citep{florou2021marginally}. Still, proton synchrotron radiation is typically used to explain  the high-energy part of the $\gamma$-ray spectrum (greater than 100 MeV) \citep{Vietri,Totani_1998,Asano_2009,Razzaque_2010}, while the sub MeV photons are produced by primary electrons that also emit synchrotron radiation \citep{dermer-attonian2003}. This scenario has been applied to explain the underlying power-law components seen in some bright Fermi-LAT bursts in the GeV energy band \citep{2008Racusin,2009Abdo}. Another scenario for the production of the high-energy part of the spectrum are proton induced cascades \citep{Dermer-atonian_2006,Asano2007}. In any case, hadronic models pose an attractive alternative to the leptonic ones for the GRB emission mechanism; they make predictions for ultra-high energy cosmic ray production \citep{Vietri_1995,Waxman_1995,Murase:2008mr} and high-energy neutrino emission \citep{waxman97b,Murase:2008sp,Gao_2012} that are testable with existing (e.g., IceCube Neutrino Telescope and future experiments, e.g. IceCube Gen2, the KM3NeT Open Science System \citep{Aartsen_2021,schnabel2021km3net}.

A disadvantage of hadronic models, when applied to  GRBs, is their low radiative efficiency, i.e. one needs very large luminosities in protons in order to produce the required GRB  $\gamma$r-ay luminosity. However, under certain circumstances, a hadronic system undergoes an abrupt transition from a radiatively inefficient to a radiatively efficient state, often exhibiting  flaring activity. This intriguing property of hadronic systems is coined as hadronic supercriticality (HSC) and many of its properties can be related to the GRB prompt emission \citep{Kazanas02,KaM,14MPGM,PM18,am20}. The relativistic protons inside the source  become supercritical once a feedback and a marginal stability criterion are simultaneously satisfied, as first demonstrated by the stability analysis of \cite{KM92}. As far as the feedback criterion is concerned, \cite{KM92} and \cite{PM12} showed that the network of processes that play a key role in the manifestation of supercriticality are synchrotron radiation, photohadronic interactions and $\gamma \gamma $ pair production. Three feedback loops are created between these physical processes that are responsible for the state transition described above. The marginally stability criterion has to do with the value of the relativistic proton column density inside the source. In case it exceeds a critical value the system behaves non linearly and bears similarities to the Lotka-Volterra type of systems that describe a prey-predator relation between protons and photons. In this project we expand the work of \cite{14MPGM,PM18,am20}, which assumed an emitting region of constant volume and magnetic field,  by investigating for the first time the same phenomenon in adiabatically expanding sources.

This paper is structured as follows: In Sec. \ref{sec2all} we investigate analytically how the supercritical behaviour can be affected by the expansion of the source. We next present  in Sec. \ref{sec2.3} our methodology and the numerical code that we utilise. In Sec. \ref{Sec:3} we investigate numerically the effect of the expansion on the manifestation of supercriticality. In Sec. \ref{Sec:4} we discuss our results in the context of GRB prompt emission. For this, we compute the spectrum and light curve of a fiducial GRB powered by several expanding blobs ejected by the central engine. Finally, we summarise and discuss our results in Sec.~\ref{sec5}. For the GRB application we adopt $z=2$ as a typical redshift and use $H_{\rm 0} = 69.32$ $\rm km \ Mpc^{-1}\ s^{-1}$, $\Omega_{\rm M} = 0.29$, $\Omega_{\rm \Lambda} = 0.71$ \citep{wmap9}. Quantities denoted with the superscript/subscript `obs' refer to the observer's frame, while the rest to the comoving frame of the outflow.

\section{First principles} 
\label{sec2all}
\label{sec2}

In the non expanding case, the HSC  manifests itself when the proton energy density inside the source exceeds some critical value -- see \cite {am20} for a comprehensive study. The above requirement poses a marginal stability criterion, according to which, above the critical proton density, the system becomes highly efficient, transferring the stored proton energy into secondary particles, namely photons, electron/positron pairs and neutrinos.  Previous work on the manifestation of supercriticality in non expanding systems, has shown that this hadronic non linear behaviour arises  as a result of specific networks of physical processes (feedback loops). For example, \cite{KM92} showed that synchrotron photons of the relativistic electron-positron pairs, produced via photo-pion/photo-pair interactions, become targets for the relativistic protons, which then produce  even more pairs and pions. This feedback loop leads to an exponential photon outgrowth and eventually to fast proton energy losses. Another type of feedback network was examined by \cite{PM12}, according to which $\gamma$-rays, that are produced directly via proton synchrotron or indirectly  via photo-pair and photo-pion interactions, turn spontaneously into electron-positron pairs and eventually in soft photons, which become targets for the relativistic protons and feedback on them.

The system enters the supercritical regime in various ways, the most interesting being with multiple photon bursts that occur quasi periodically (limit cycles) when the number density of protons slightly exceeds the critical value. For even higher proton densities, the number of bursts in the light curve is increased. However, if the proton number density keeps increasing the temporal behaviour of the system degenerates to a single burst before it saturates into a  highly efficient steady state. As it was shown in \cite{am20}, the critical proton density depends on several source parameters like the radius of the source, the magnetic field strength and the specifics of the proton distribution (minimum/maximum proton Lorentz factors and the power law slope assuming that it is a power law). The main question that we are going to address in the present project is whether the supercritical behaviour persists when the source is spherically expanding and (if yes) how it manifests itself.

We begin our analysis by assuming an expanding spherical volume of instantaneous radius:
\begin{equation}
    r=r_{\rm in}+u_{\rm exp}(t-t_{\rm in}).
    \label{eq:rt}
\end{equation}
Here $u_{\rm exp}$  is the expansion velocity which we assume to be constant, $t$ is the time, $t_{\rm in}$ is the initial moment of the particle injection inside the source volume and $r_{\rm in}$ the initial radius of the source.

In the case of no particle escape and negligible energy losses, the differential equation which describes the evolution of the number density of an energy integrated proton distribution inside the source, is given by:
\begin{equation}
u_{\rm exp} \frac{d n_{\rm p}}{d r}+\frac{3 u_{\rm exp} n_{\rm p}}{r}=\mathcal{Q}_{\rm p},
\label{keqap}
\end{equation}
where $\mathcal{Q}_{\rm p}$ is the proton injection rate per unit volume:
\begin{equation}
    \mathcal{Q}_{\rm p}(r)=\mathcal{Q}_{\rm p,0} \left(\frac{r}{r_{\rm in }}\right)^{s-3},
\end{equation}
 $\mathcal{Q}_{\rm p,0}$ is a normalisation factor and  $s$ is an index that determines the rate of energy injection as the source expands. We take as an initial condition that $n_{\rm p}(r_{\rm in})=0$. The solution  of eq. \ref{keqap}, for $s \neq -1$, is given by the following relation:
\begin{equation}\label{np}
  n_{\rm p}(r)=\frac{\mathcal{Q}_{\rm p,0} r^{s-2}}{u_{\rm exp} (s+1)}-\frac{\mathcal{Q}_{\rm p,0}r_{\rm in}^{s+1}}{u_{\rm exp} (s+1) r^{3}},
  \end{equation}
while in the case $s=-1$, the solution is:
\begin{equation}
\label{np2}
    n_{\rm p}(r)=\frac{\mathcal{Q}_{\rm p,0} \ln(r/r_{\rm in})}{u_{\rm exp} r^{3}}.
\end{equation}
As it can be seen from the above relations, since the initial condition is $n_{\rm p}(t_{\rm in})=0$, at early times proton injection will cause  an increase in the number density, which will reach a maximum, for the cases where $s<2$, and then will drop for $r>>r_{\rm in}$. For $s>2$ the proton number density will keep increasing, while in the case of $s=2$, it will reach asymptotically a constant value.

\begin{figure}
    \centering
    \includegraphics[width=0.9\columnwidth]{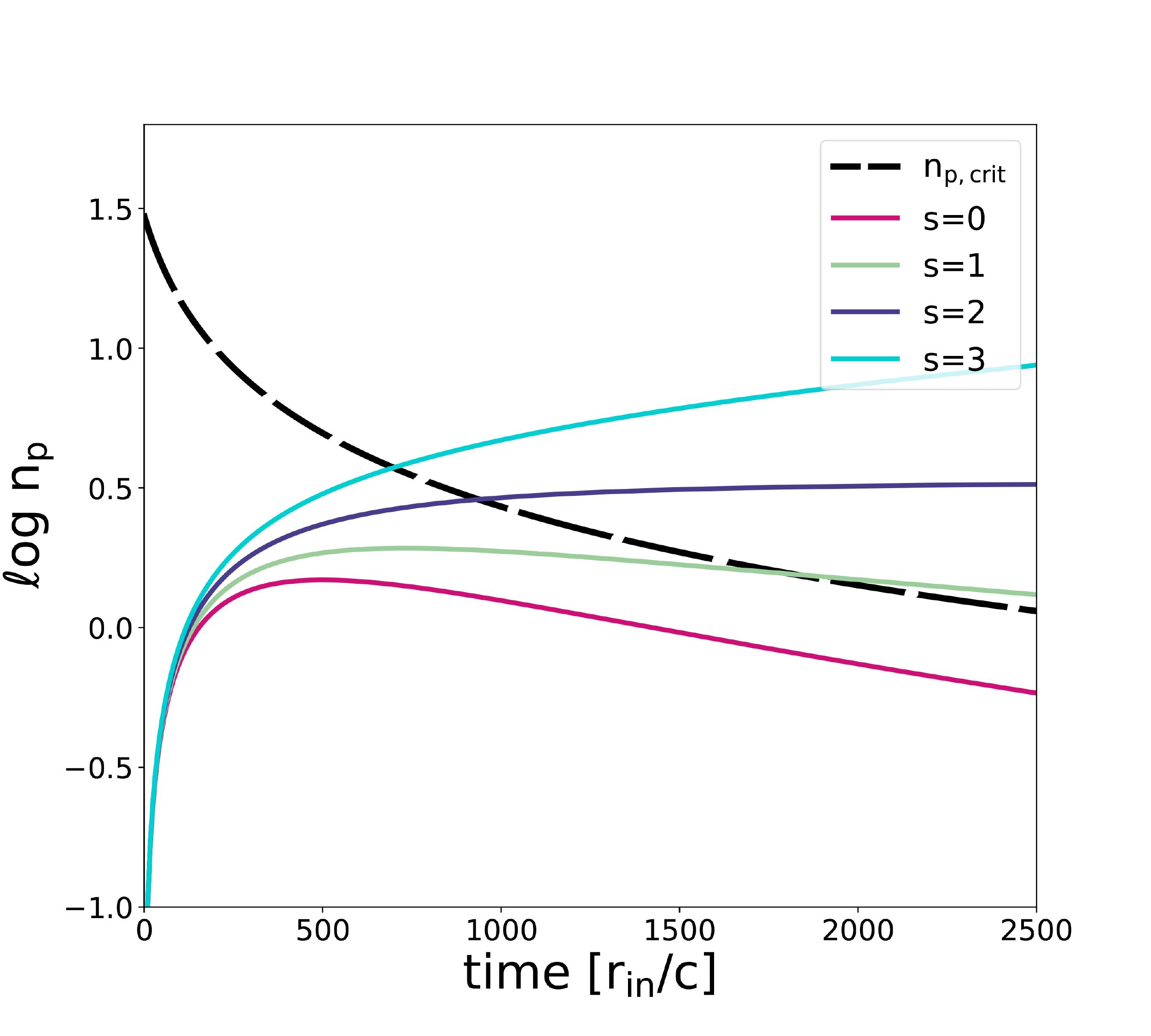}
        
    \caption{The proton number density derived from eq. \ref{np}, in arbitrary units, as a function of time, expressed  in $r_{\rm in}/c$ units, for a fixed value of the expansion velocity, $u_{\rm exp}=10^{-3}c$, and different values of the proton luminosity injection index $s$. The dashed black line corresponds to the marginal stability criterion for the PeS loop (see eq. \ref{marginalst}).}
    \label{fig:toyexamp}
\end{figure}

Figure~\ref{fig:toyexamp} illustrates the evolution of the proton number density (see eq. \ref{np}) with time (in $r_{\rm in}/c$ units), for different values of the proton luminosity injection index $s$ (coloured lines). As an example, we also plot, with a black dashed line, the critical proton density required to reach the marginal stability criterion for the PeS feedback loop \citep{KM92}. This is formed when the synchrotron photons radiated from the photo-pair secondaries produce even more pairs on the protons before they escape from the source. The critical number density $n_{\rm p,crit}$, for a proton Lorentz factor $\gamma$, is expressed according to the following relation:
\begin{equation}
 n_{\rm p,crit}=  \left(\frac{2}{3}p-1 \right) \ b^{1-p/3} \ \left(\int_{2}^{b\gamma^{3}} \sigma_{\rm pe}(y) y^{-1-p/3} dy\right)^{-1} r^{-1}. 
 \label{marginalst}
\end{equation}
This expression has been derived for a proton power-law distribution with index $p$. In the example shown in Fig.~\ref{fig:toyexamp} we used $p=2$. The lower limit of the integral is the threshold
of the photo-pair interaction (in units of $m_{\rm e} c^2$), while the upper limit is the relation that defines the critical value of proton Lorentz factor  for which the PeS feedback loop operates, i.e. $b\gamma^{3}  \ge 2$. Here $b=B/B_{\rm crit}$, $B$ the magnetic field inside the source, while $B_{\rm crit}=4.4\times 10^{13}$ G is the Schwinger magnetic field and  $\sigma_{\rm pe}$  is the cross section of the photo-pair interaction. For the specific example shown in  Fig.~\ref{fig:toyexamp} we have used 
$B=10^{4}$ G.

For all practical purposes, we assume that when the curve critical proton density intersects the curves of the proton densities, then the system enters the supercritical regime. 
The critical proton density decreases as $1/r$ (see eq.~\ref{marginalst}) as the source expands (assuming a constant magnetic field). Therefore it will always intersect the curves with $s \geq 2$, while it might or it might not intersect the ones for $s<2$, depending on $Q_{\rm p,0}$. This is illustrated in Fig. \ref{fig:toyexamp} where $n_{\rm p,crit}$ (black dashed line) intersects the proton number density for $s=1$ (green line) but not for $s=0$ (pink line). In this latter case one would have to increase the value of $Q_{\rm p,0}$ in order to bring the system in the non linear regime. Therefore, the onset of supercriticality becomes more luminosity demanding, as the value of proton luminosity injection index becomes smaller. Furthermore, inspection of eqs. \ref{np} and \ref{np2} reveals that when the velocity of expansion becomes higher and $s<2$, the system can become supercritical only for higher values of the proton injection rate. 

If one integrates over time the coloured curves depicted in Fig. \ref{fig:toyexamp} until the moment each one intersects the black dashed line, the results are similar in all cases.  In other words, even though the proton injection rates needed for the onset of a supercritical flare are higher  for $s<2$ compared to those for $s\ge2$, the proton column density needed for supercriticality is independent of $s$. This is reminiscent of nuclear piles \citep{Kazanas02,KaM}.

\section{The Numerical Code}
\label{sec2.3}
In order to study the full problem, taking into account all the radiative processes, we have to solve the proton, photon and electron/positron kinetic equations numerically. We consider, as previously, a spherical source of initial radius $r_{\rm in}$ that expands adiabatically with a constant velocity $u_{\rm exp}$.  The emitting region contains a tangled magnetic field of strength $B$, which varies with $r$ as the source expands according to:
\begin{equation}
   B=B_{\rm in} \left(\frac{r_{\rm in}}{r} \right)^{q} 
\end{equation}
where $B_{\rm in}$ is the value of the magnetic field at $r=r_{\rm in}$.

We assume that pre-accelerated relativistic protons are injected in the source, having a power-law energy distribution with a luminosity depending on the location of the emitting region in the jet, expressed as:
\begin{equation}
  L_{\rm p}(r)=L_{\rm p,in} \left( \frac{r} {r_{\rm in}} \right)^s 
  \label{Lp}
\end{equation}
where $L_{\rm p,in}$ is the proton injection luminosity at the initial radius of the source.
The proton injection luminosity is related to the differential proton injection rate, which enters the equations we have to solve, as:
\begin{equation}
Q_{\rm p}(\gamma,r)=\frac{L_{\rm p}(r)}{(m_{\rm p}c^{2})^{2} \int^{\gamma_{\rm max}}_{\gamma_{\rm min}} \gamma^{-p+1} d\gamma},
\label{eq:qp}
\end{equation}
where $Q_{\rm p}(\gamma,r)$ is defined as:
\begin{equation}
Q_{\rm p}(\gamma,r)\equiv \frac{dN_{\rm p}}{d\gamma dr}=Q_{\rm i}(r)\gamma^{-p}H(\gamma-\gamma_{\rm min})H(\gamma_{\rm max}-\gamma) H(r-r_{\rm in})
\label{eq:injection}
\end{equation}
and the proton injection per volume: 
\begin{equation}
    \mathcal{Q}_{\rm p}(\gamma,r)=\frac{3 Q_{\rm p}(\gamma,r)}{4\pi r^{3}}
    \label{eq:injection2}
\end{equation}
For the definition of $\mathcal{Q}_{\rm p}(\gamma,r)$ see the previous section. In eq. \ref{eq:injection} $H(x)$ is the Heaviside step function, $\gamma_{\rm min}$, $\gamma_{\rm max}$ are the minimum and maximum proton Lorentz factors, respectively, and $Q_{\rm i}$ is  a radially dependent normalisation factor. Since we assume that $\gamma_{\rm min}$ and $\gamma_{\rm max}$ remain constant throughout the evolution of the system, the above relations imply that $Q_{\rm i}\propto r^{-s}$. 

It is also useful to define  a measure of the proton luminosity in terms of the proton compactness:
\begin{equation}
\ell_{\rm p} = \frac{\sigma_{\rm T} L_{\rm p}}{4 \pi r m_{\rm p} c^{3}} = \ell_{\rm p,in} \left(\frac{r}{r_{\rm in}}\right)^{s-1}
\label{eq:lpin}
\end{equation}
Similarly, we define the instantaneous photon compactness $\ell_{\rm \gamma}$:
\begin{equation}
\ell_{\rm \gamma}=\frac{ \sigma_{\rm T} L_{\rm \gamma} }{4 \pi r m_{\rm e} c^{3}}
\label{eq:lg}
\end{equation}
where $L_{\rm \gamma}$ is the bolometric photon luminosity of the source.

Protons, electrons and photons are the three stable species inside the source. We assume that pions and muons decay instantaneously, neutrons do not interact with soft photons, and neutrinos escape the source freely. 
The evolution of the stable particle populations inside the spherical volume can be described by a system of coupled integro-differential kinetic equations: 
\begin{equation}
 \frac{\partial n_{\rm j}}{\partial t}+ u_{\rm exp} \frac{3 n_{\rm j} }{r(t) }+ \frac{n_{\rm j}}{t_{\rm esc,j}}+\mathcal{L}_{\rm j}= \mathcal{Q}_{\rm j}
\label{kineticgeneral}
\end{equation}
where the index $j$ refers to protons (denoted as p), electrons/positrons (denoted as e) and photons (denoted as $\gamma$), $t_{\rm esc,j}$ is the escape timescale from the source, and  $n_{\rm j}$ is the differential number density of each species. We assume that all charged particles remain confined in the blob ($t_{\rm esc, p/e} \rightarrow \infty$) and only photons escape on a timescale $t_{\rm esc,\gamma}=r/c$ (same for neutrons and neutrinos). The loss ($\mathcal{L}_{\rm j}$) and injection ($\mathcal{Q}_{\rm j}$) per volume terms\footnote{For protons the injection operator is given by eq.~\ref{eq:injection2}, and is equal to zero for relativistic primary electrons.} include the following processes (for details see \citealt{95km}): 
\begin{itemize}
   \item synchrotron radiation for both electrons and protons
    \item proton-photon pair production (photopair)
    \item proton-photon pion production (photopion)
    \item synchrotron self-absorption
    \item electron inverse Compton scattering
    \item photon-photon ($\gamma \gamma$) pair production
    \item electron-positron pair annihilation
    \item adiabatic losses, as described in  \cite{kardashev}.
\end{itemize}

We develop a new numerical code starting from the one of \cite{95km,97km} that was applicable to non expanding sources. This new version solves the system of coupled integrodifferential described by eq.~\ref{kineticgeneral}, and gives the evolution of the distribution of the stable particle populations as a function of the continuously changing comoving source radius. We note that we treat the synchrotron and inverse Compton scattering as full emissivities while we take delta-function approximations for the photopion process. A first version of this code which contains only leptonic processes has been presented recently in \cite{BoulaMasti} in application to the non-thermal emission from AGN.

The free parameters of the problem are the initial radius $r_{\rm in}$ and magnetic field strength $B_{\rm in}$ of the source, the power-law index of the magnetic field radial profile $q$, the initial proton compactness $\ell_{\rm p,in}$, the power-law index of the proton injection radial profile $s$, the maximum proton Lorentz factor $\gamma_{\rm max}$ and the expansion velocity $u_{\rm exp}$. 
We assume that $\gamma_{\rm min}=1$ and $p=2$, and we take as an initial condition that  $n_{\rm j}(\gamma_{\rm j},r_{\rm in})=0$.

\section{HSC in expanding sources}
\label{Sec:3}

\begin{figure}
  \includegraphics[width=\columnwidth]{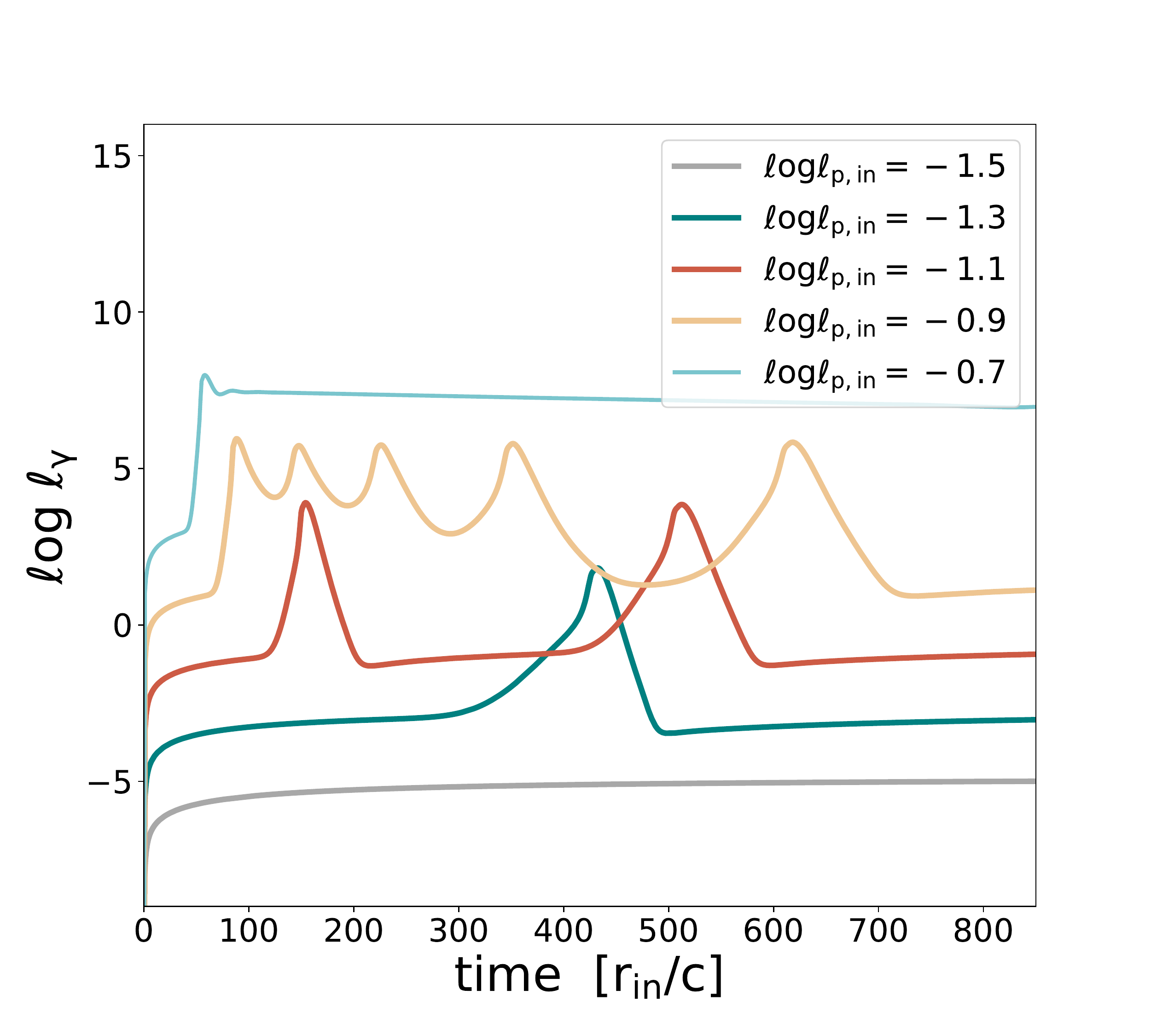}
    \includegraphics[width=\columnwidth]{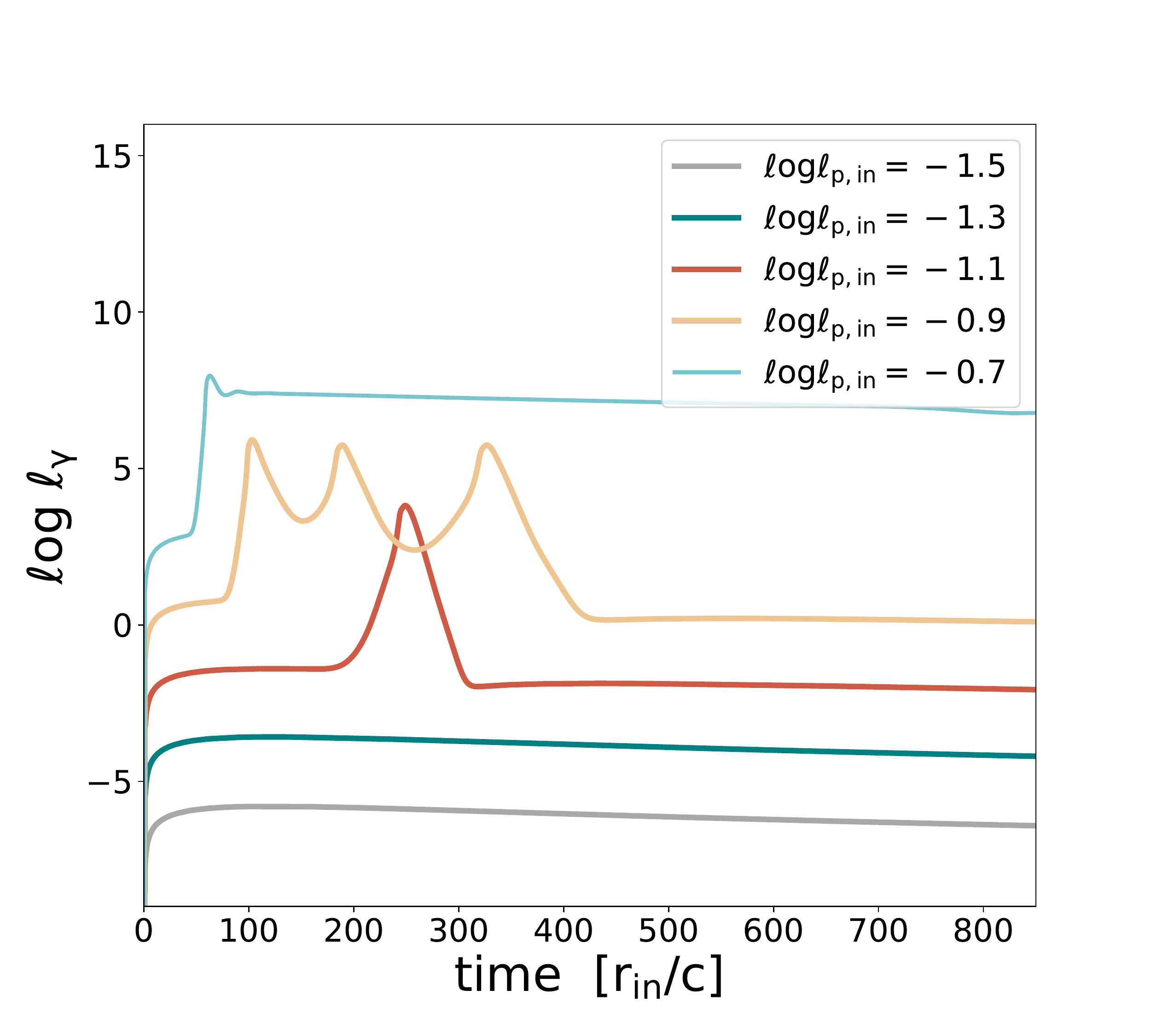}
\caption{Top panel: logarithmic plot of photon compactness as a function of time in $r_{\rm in}/c$ units, for various values of the initial proton injection compactness that differ by a factor of 0.2 in logarithm. The source  magnetic field is constant ($q=0$). For better readability, 
each curve is shifted from the previous one by a factor of 2. Bottom panel: Same as in the top panel but for a decaying magnetic field 
($q=1$). In both panels the source is expanding with velocity $u_{\rm exp}=10^{-2.5}c$. Other parameters used are: $B_{\rm in}=10^{4}$ G,  $r_{\rm in}=10^{11}$ cm, $\gamma_{\rm max}=10^{4}$ and $s=0$. }
  \label{Fig1}
\end{figure}

In this section we examine numerically the effects of the expansion on the onset and phenomenology of HCS.

\subsection{Comparison to non expanding sources }

\label{Sec:3.1}

We begin the numerical investigation by examining the modification  that expansion would bring to the onset of HSC. To have the closest possible analogy with the non expanding case, we neglect adiabatic losses while we assume that the magnetic field and proton injection luminosity do not change with radius,  i.e.  $q=0$ and $s=0$ respectively. We then select a low value for the expansion velocity, $u_{\rm exp}=10^{-2.5}c$
 and  use a set of parameter values that would drive a non expanding source to supercriticality, namely $\ell_{\rm p,in} = 10^{-3}$, $r_{\rm in}=10^{11}$~cm, $\gamma_{\rm max}=10^{4}$, and $B_{\rm in}=10^{4}$~G. Nevertheless, when the system expands the numerical result is a subcritical light curve, as  depicted in grey on the top panel of Fig.~\ref{Fig1}. We start increasing  $\ell_{\rm p,in}$ until we find a value that produces one supercritical flare (teal solid curve on the top panel of Fig. \ref{Fig1}). This value, at least within the accuracy of our numerical resolution, marks the onset of the HSC in this case and we denote it  as $\ell_{\rm p,crit}$.  From there on, we progressively increase $\ell_{\rm p,in}$, by a factor of 0.2 in logarithmic scale and record the system's response.  As in the non expanding case, multiple outbursts occur more frequently as the compactness increases, while the first flare in each case progressively appears at earlier times. For even higher proton compactnesses, the system enters supercriticality but saturates very quickly with the photon compactness reaching a constant value (light blue solid curve on the top panel of Fig. \ref{Fig1}).

Overall, our results are analogous to those of previous works on HSC  for non expanding systems. Nonetheless, there are two main differences. First, the non linear behaviour in an expanding source is achieved at the cost of higher proton luminosities; a higher injection rate of protons is needed to counterbalance the effects of expansion and to reach the critical number density (see Sec.~\ref{sec2all}). Second, the duration of flares in the supercritical regime increases with time as a result of the source expansion (see, e.g. the red and yellow coloured light curves in Fig. \ref{Fig1}).

We can summarise the phenomenology of the system with increasing proton luminosity as follows: subcritical steady state $\xrightarrow[]{}$ one supercritical flare $\xrightarrow[]{}$ multiple supercritical flares $\xrightarrow[]{}$  a single flare that quickly merges into a supercritical steady state.

\subsection{The effect of the magnetic field radial profile}

We now extend our analysis by examining the more realistic case where the magnetic field is decreasing as the source volume increases  -- see Fig \ref{Fig1}, bottom panel. Choosing exactly the same initial conditions that produced the light curves on the top panel of  Fig. \ref{Fig1} but letting $B$ to vary as $r^{-1}$ we find that one needs to increase the proton luminosity for the system to enter the supercritical regime --  compare teal and red lines in the top and bottom panel of Fig. \ref{Fig1}. 
Even when well inside the supercritical regime, the number of bursts tends to decrease when $B$ decreases with distance -- compare the orange lines in the top and bottom panels of Fig. \ref{Fig1},
making, thus, the HSC less efficient. As expected, this trend is intensified when B drops faster with distance.
We therefore conclude that the decrease of the magnetic field as the source expands tends to suppress the HSC. Note, however, that its basic characteristic features are still there.

\subsection{The effect of expansion velocity }
\label{sec4.3}

\begin{figure}
    \centering
    \includegraphics[width=\columnwidth]{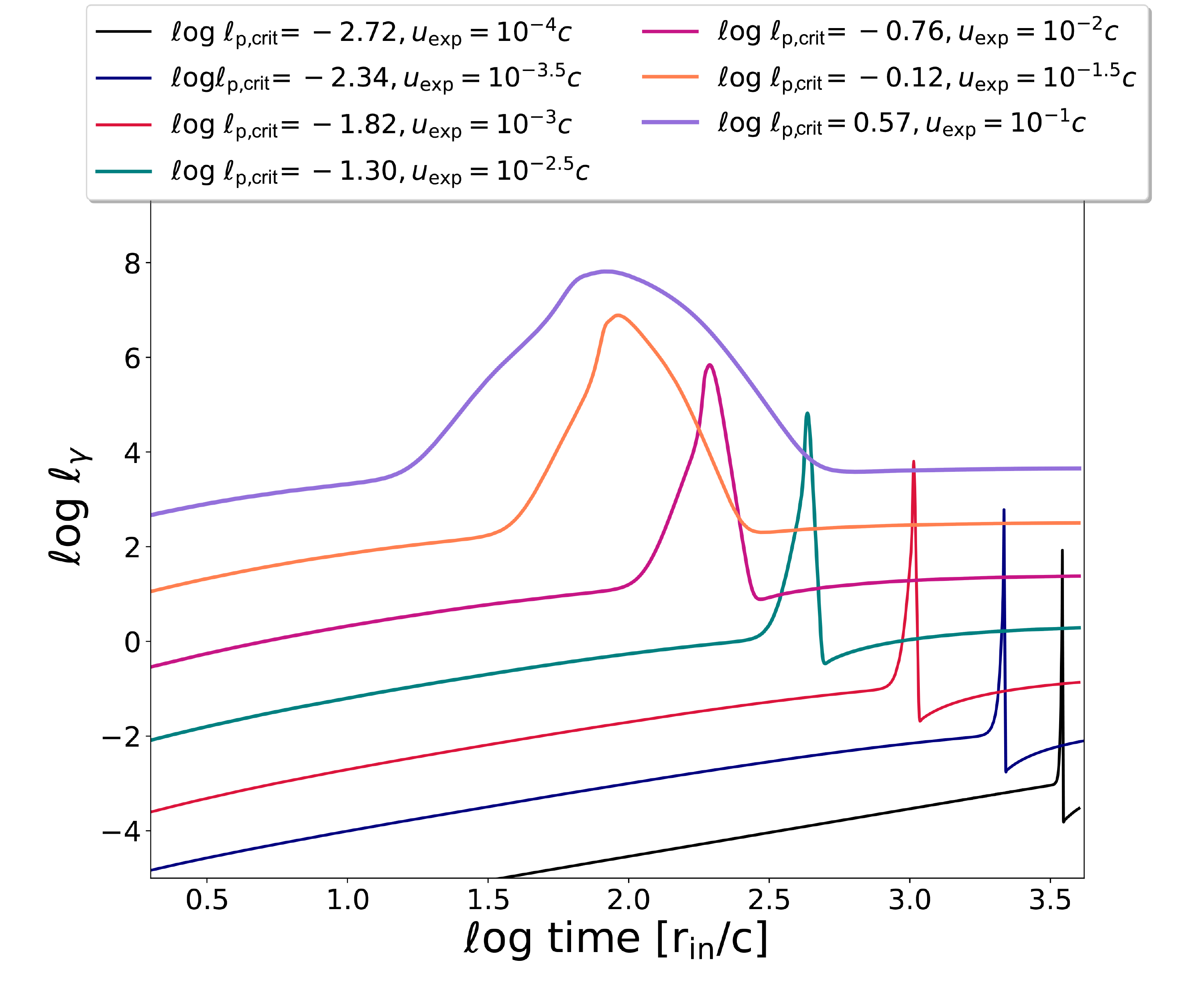}
    \caption{ A log-log plot of photon compactness as a function of time in $r_{\rm in}/c$ units, for different values of $u_{\rm exp}$ and $\ell _{\rm p,crit}$.  Each light curve is shifted from the previous one in the y-axis one order of magnitude to make the plot easier to read. Other parameters are: $r_{\rm in}=10^{11}$ cm, $B_{\rm in}=10^{4}$ G, $\gamma_{\rm max}=10^{4}$, $q=0$, $s=0$. }
    \label{fig:2}
\end{figure}

We investigate next how the supercritical behaviour is affected by the expansion velocity. We choose the same set of parameters as the ones used for the top panel of Fig. \ref{Fig1}. Initially we choose a very low  value for expansion velocity in order to have an estimate of the system's behaviour in the limiting case of no expansion. For $u_{\rm exp} = 10^{-4}c$ the first supercritical flare appears at $4000 \ r_{\rm in}/c$ (black solid line in Fig. \ref{fig:2}). This is the minimum time that is needed for producing at least one photon outburst (henceforth, $T_{\rm B}$), and is a function of model parameters, i.e. $T_{\rm B}=f(B_{\rm in}, \gamma_{\rm max},u_{\rm exp},s,q)$. If we start  increasing the expansion velocity, e.g. $u_{\rm exp}=10^{-3.5}c$, while keeping all other parameters the same, the supercritical behaviour is lost. This is due to the fact that the increase of the expansion velocity  leads to a decrease of the proton density  and as a result the marginal stability criterion is never met. In order to recapture this condition one needs to increase the injected proton luminosity. In this case the system shows a flare earlier than before (see blue light curve in Fig. \ref{fig:2}). This procedure is repeated for greater values of the expansion velocity (see coloured light curves of Fig. \ref{fig:2}) and a similar behaviour is found, i.e. the larger the value of expansion velocity, the earlier the appearance of the first supercritical flare.
This occurs, however, at the expense of more proton luminosity. These results are in total agreement with our analysis in Sec.~\ref{sec2all}.

We then explore the temporal behaviour of the system for different values of the expansion velocity following the procedure outlined in Sec. \ref{Sec:3.1}. For each light curve depicted in Fig.~\ref{fig:2} we increase the value of $\ell_{\rm p,in}$ and record the shape of the produced light curve. We find  that as the expansion velocity increases, the number of multiple bursts decreases and finally, for $u_{\rm exp}>10^{-2}c$, it reduces to a single burst (not explicitly shown here). This phenomenology is similar to the one shown in Fig.~\ref{Fig1}. We also note that the increase of the expansion velocity leads to the production of broader 
photon pulses. The width of the pulse is related to the light crossing time of the emitting source, which is larger for higher values of $u_{\rm exp}$ at a given time. Moreover, the system's radiative efficiency, defined as the ratio of the total
radiated energy in photons to the total injected energy in relativistic protons within a constant time interval, decreases as the source expands faster.

\subsection{The effects of adiabatic losses and of the proton luminosity profile}
\begin{figure}
\centering
\includegraphics[width=\columnwidth]{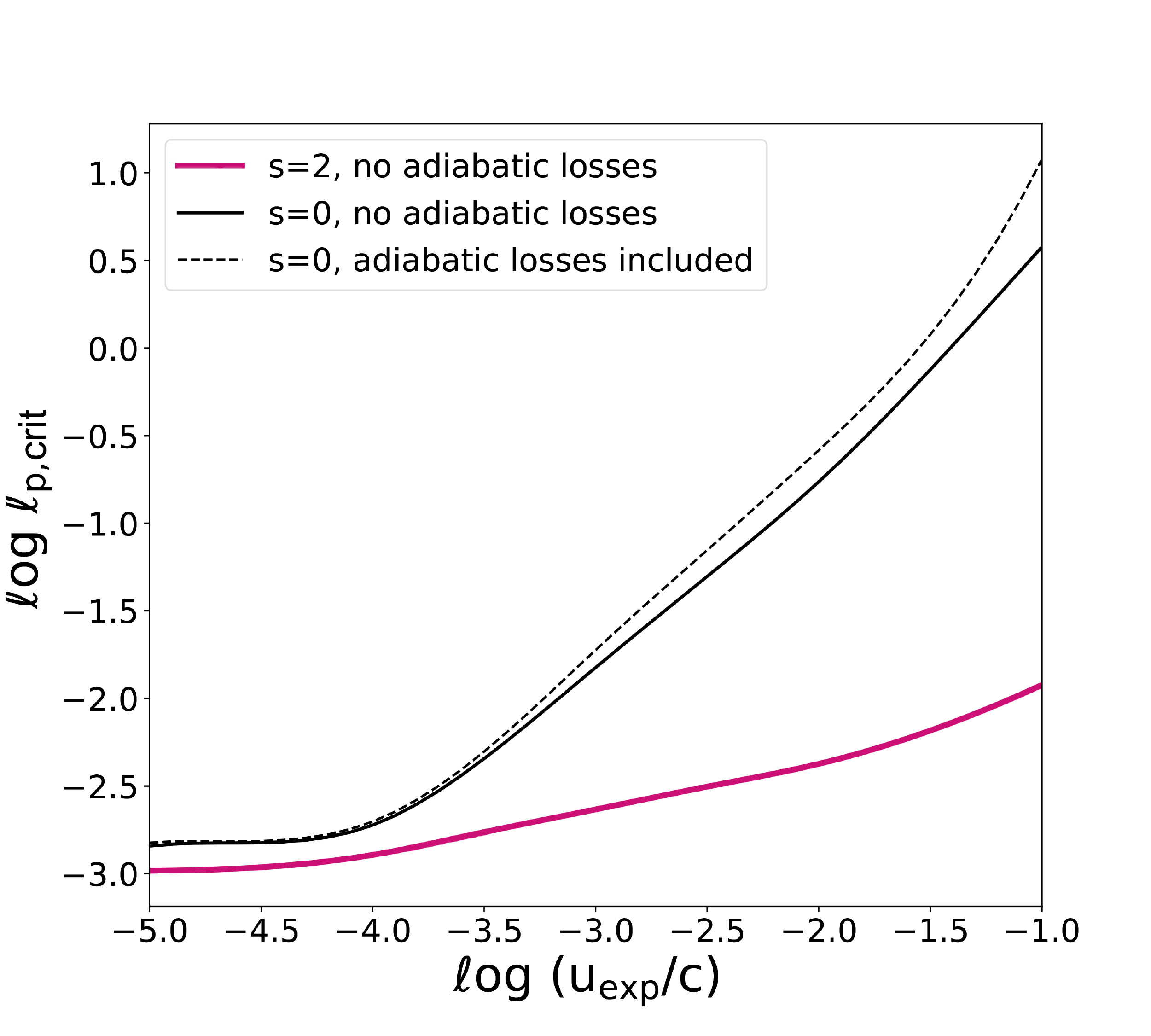}
    \caption{Plot of $\ell_{\rm p,crit}$ as a function of $u_{\rm exp}$ for $s=0$ without adiabatic losses for particles included (black solid line), $s=0$ with adiabatic losses 
    of particles included
    (dashed black line), and $s=2$ (magenta solid line). All other parameters are the same as in Fig. \ref{fig:2}. }
   \label{fig:4lpuexp}
\end{figure}

All results presented so far were derived without taking into account the adiabatic losses in the kinetic equations of protons and electrons. These are not expected to be important for most parameters studied here, except for the lowest energy particles. We therefore repeated the calculations shown in Fig. \ref{fig:2} after including adiabatic energy losses and record the  $\ell_{\rm p, crit}$ that is needed to bring the system to supercriticality. The results are plotted as a function of $u_{\rm exp}$ in Fig. \ref{fig:4lpuexp} (dashed black line) and should be compared with those obtained without adiabatic losses (solid black line). The difference in the required proton injection compactness between the two cases is negligible, except for high expansion velocities where differences up to a factor of $\sim 3$ are found.

Another parameter that plays a role in the appearance of supercriticality is the radial profile of the proton luminosity. In all the above we have assumed that the proton luminosity is independent of the source radius (and location along the jet). We therefore repeated the calculations used for the construction of Fig. \ref{fig:2} with the same parameters, except for the injected proton luminosity index $s$. As we have already demonstrated with a simple model in Sec. \ref{sec2}, the increase of $s$ makes the appearance of supercriticality less luminosity demanding. This is demonstrated in Fig.~\ref{fig:4lpuexp} where $\ell_{\rm p,crit}$ is plotted for $s=2$ as a function of expansion velocity (solid magenta line). The difference with the case where $s=0$ is that for $s\ge 2$ one can always find one supercritical flare for a fixed proton compactness value, even when the source expands with a fast velocity. In the case of very slow expansion, e.g. $u_{\rm exp}=10^{-4}c$, a supercritical flare is produced at  $T_{\rm B}=4000\ r_{\rm in}/c$. As we have already mentioned, this value depends on the set of  chosen source parameters. For higher expansion velocity values and the same $\ell_{\rm p,crit}$ this supercritical flare is found at later moments. We therefore increase the $\ell_{\rm p,in}$ in order to fix the appearance of the flare at $T_{\rm B}=4000\ r_{\rm in}/c$. We conclude that the proton compactness is lower than the corresponding value  for $s=0$. However, the amount of proton energy required for the system to become supercritical is independent of $s$, in agreement with our analytical findings  (see Sec. \ref{sec2all}).

\section{Relevance to GRB prompt emission}
\label{Sec:4}

\subsection{Energetics and photon spectra }
\label{sec3.3}

In this section we discuss HSC  in the context of GRB prompt emission. We assume that at a distance $R_{\gamma}$ from the central engine protons are accelerated to a power law  and are subsequently injected into a spherical region with radius $r=R_{\gamma}/ \Gamma$ as measured in the jet comoving frame. The assumption of the spherical geometry is valid as long as the beaming angle $1/\Gamma$ is smaller than the opening angle of the GRB jet. This spherical region moves away from the central engine with a bulk Lorentz factor $\Gamma$ and is expanding with a velocity $u_{\rm exp}$.

We fix $r_{\rm in}=10^{11}$ cm,  $B_{\rm in}=10^{4}$ G,  which can be considered as nominal values for GRBs, and run the code for different values of the maximum proton Lorentz factors in the range $10^{4}-10^{8}$. We numerically verified that for $\gamma_{\rm max} \le 10^{4}$ the multi-burst behaviour is lost. We therefore exclude these Lorentz factors from the analysis. As an indicative example, we take the magnetic field of the emitting region to drop linearly with radius (i.e. $q=1$) as the source is expanding, and the proton injection luminosity to increase as $s=1$. Finally, we assume that the shell expands adiabatically having an expansion velocity equal to $u_{\rm exp}=10^{-2.5}c$. For the transformation of quantities from the comoving to the observer's frame we use $\Gamma=100$.

\begin{figure}
    \centering
    \includegraphics[width=\columnwidth]{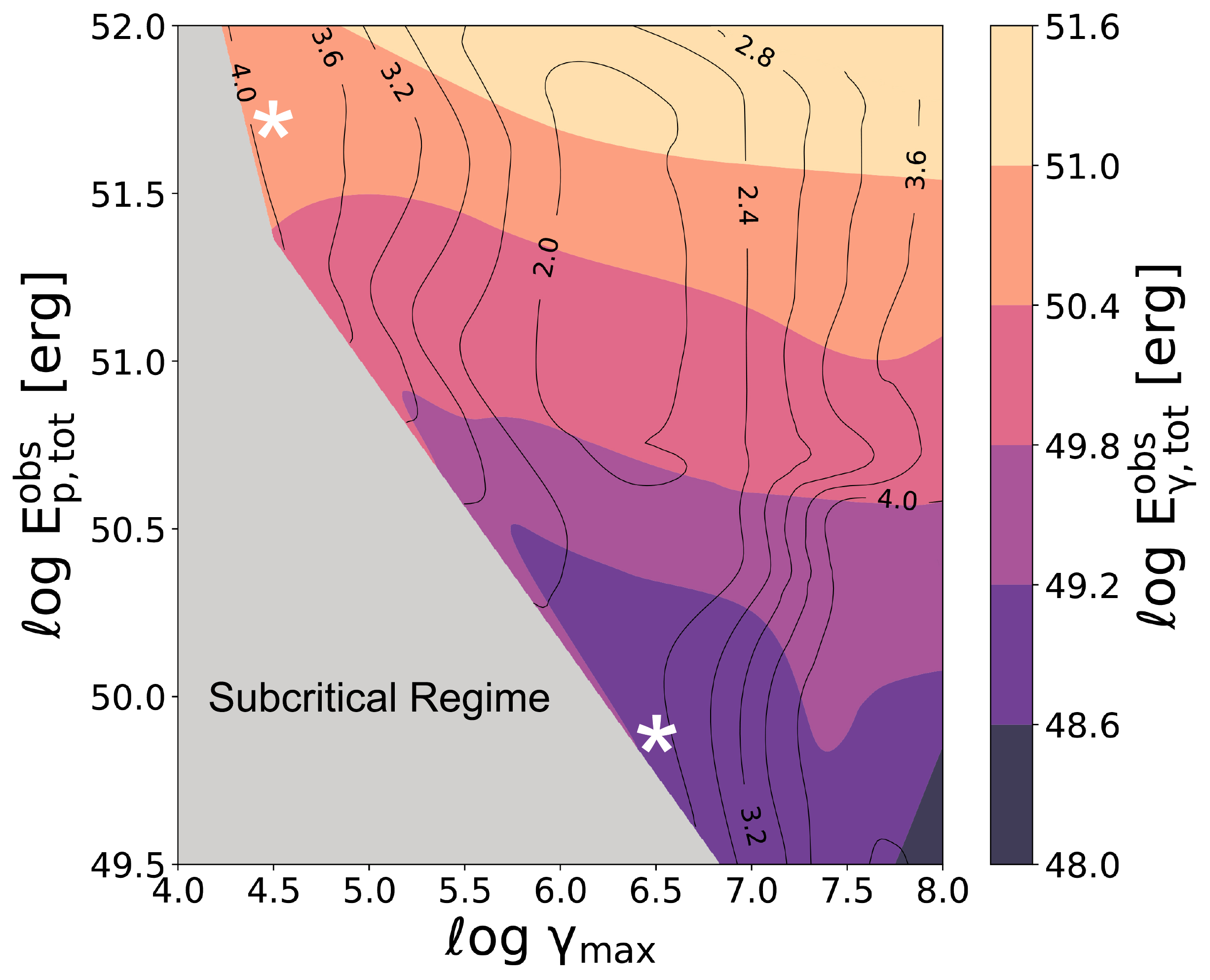}
    \caption{Phase space of the total proton energy injected in the expanding source for a time period of $T_{\rm B}=400 \, r_{\rm in}/c$ as a function of the maximum proton Lorentz factor. The coloured region corresponds to the supercritical regime while the grey space to the linear subcritical regime. Colours represent the bolometric photon energy released, while contours show the ratio of the proton density to magnetic field density $U_{\rm p,pk}/U_{\rm B,pk}$ at the peak of the first flare. The model parameters are $r_{\rm in}=10^{11}$ cm, $B_{\rm in}=10^{4}$ G, $q=1$, $s=1$, $u_{\rm exp}=10^{-2.5}$c, $\Gamma=100$ and $z=2$. White asterisks indicate two cases whose spectra and light curves are displayed in Fig. \ref{fig:contourplot}.}
\label{fig:contourplot}
\end{figure}

We inject relativistic protons in the source until 
$T_{\rm B}$, which as we discussed in Sec. \ref{sec4.3}, is  the minimum time  that is needed for producing at least one photon outburst, depending of the chosen set of parameters.
For the adopted parameter values, we find  that $T_{\rm B}\approx 400 \,  r_{\rm in}/c$ in the case where $u_{\rm exp}=10^{-2.5}c$ and $s=1$. After searching for the minimum initial proton compactness (for a given $\gamma_{\max}$) required for the onset of supercriticality, we perform consecutive runs by increasing this value by a factor of  0.1 in logarithmic scale. For each case we compute the total energy in relativistic protons (in the observer's frame) injected in the time interval between $t_{\rm in}=0$ and $T_{\rm B}$:
\begin{equation}
    E_{\rm p,tot}^{\rm obs}=\Gamma^{3} (1+z) \int_{t_{\rm in}}^{T_{\rm B}} L_{\rm p}(t) { \rm d}t.
\end{equation}
where $L_{\rm p}(t)$ is calculated from eq. \ref{eq:lpin} and time $t$ is related to the radius through eq.~(\ref{eq:rt}).
We also calculate the bolometric photon energy (in the observer's frame) released during
the same time interval:
\begin{equation}
    E_{\rm \gamma,tot}^{\rm obs}=\Gamma^{3} (1+z) \int_{t_{\rm in}}^{T_{\rm B}}  L_{\rm \gamma}(t) {\rm d}t. 
    \label{bolenergy}
\end{equation}
where $L_{\rm \gamma}(t)$ is calculated by eq. \ref{eq:lg}. We then construct a two-dimensional plot of the total proton energy used as a function of the maximum proton Lorentz factor and show the result in Fig. \ref{fig:contourplot}. The coloured region corresponds to the supercritical regime, with colour indicating $E_{\rm \gamma, tot}^{\rm obs}$, while the grey shaded region indicates the subcritical regime. For the smaller and larger values of the maximum proton Lorentz factor that we explored, i.e. $\gamma_{\rm max}=10^{4}-10^{5}$ and  $\gamma_{\rm max}=10^{6.5}-10^{8}$ respectively, the light curves exhibit multiple bursts for most values of the proton initial compactness. Only when $\ell_{\rm p,in}$ surpasses a certain value the multiple bursts blend into a single burst, as it was shown in Fig. \ref{Fig1}.  For intermediate values of the maximum proton Lorentz factor, however, i.e. $\gamma_{\rm max}=10^{5.5}-10^{6}$ we find only single-pulse supercritical light curves. This behaviour is similar to the one found for the non-expanding source 
by \citet{am20}.

\begin{figure*}
    \centering
    \includegraphics[width=1\columnwidth]{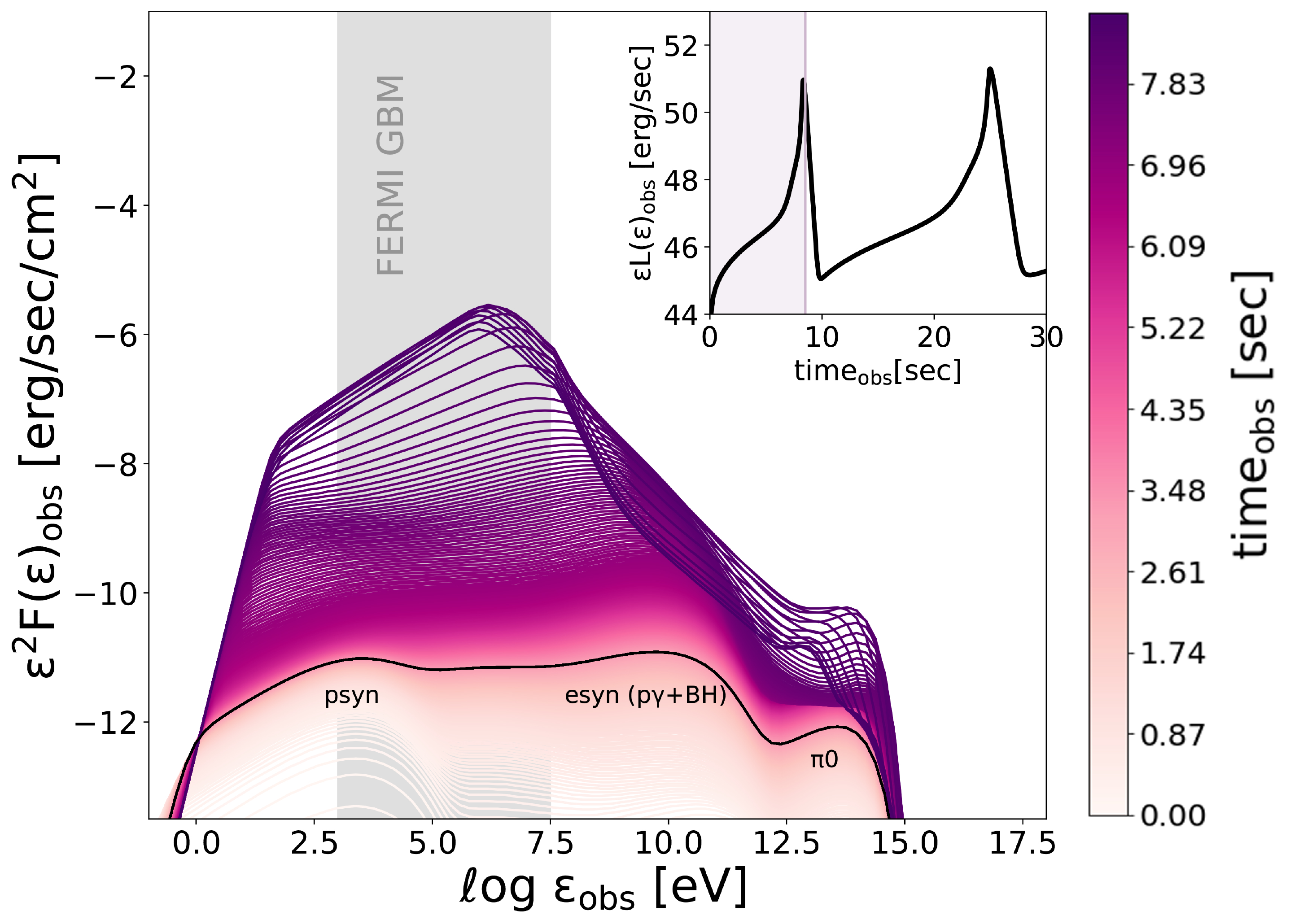}
       \includegraphics[width=1.02\columnwidth]{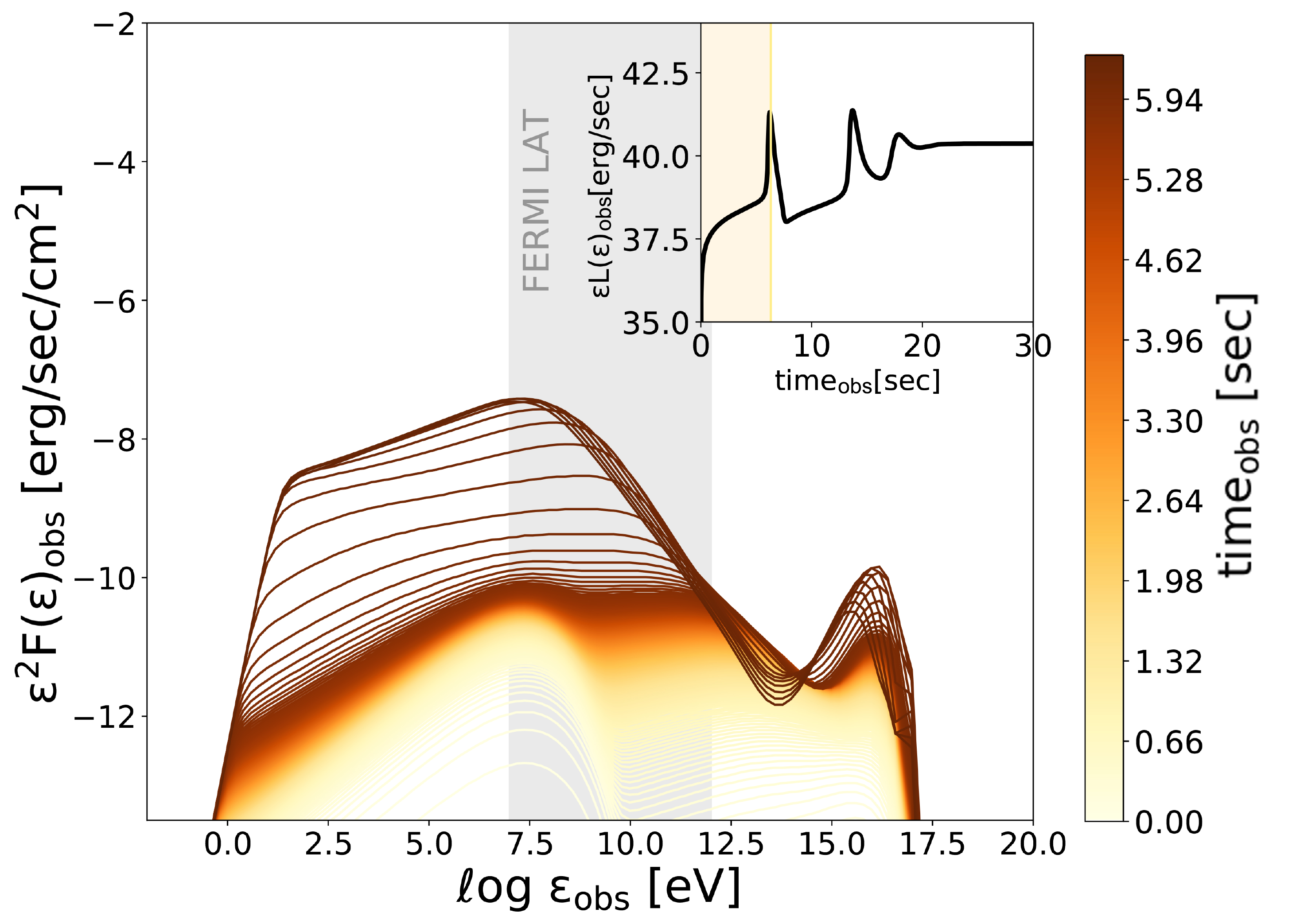}
        \caption{Left panel: A supercritical case chosen from the runs of Fig. \ref{fig:contourplot} in the case where $\gamma_{\rm max}=10^{4.5}$ (shown in the inset plot) and the corresponding photon spectra (shown in the main plots) . The  spectra correspond to the time period of the outburst's rise. This time period is also shown in highlighted stripe in the inset plot. Each spectrum differs from the previous one 1 $r_{\rm in}/c$ and this is also shown in the colour bar. Right panel: Same as in the left panel for a supercritical case where $\gamma_{\rm max}=10^{6.5}$ chosen from   the runs of Fig. \ref{fig:contourplot}. In both cases we have not taken into account the photon attenuation by EBL for the spectra construction. The vertical grey regions depict the energy bands  that the Fermi LAT and Fermi GBM observe.  The above spectra are computed in the observer's frame, assuming $\Gamma=100$ and $z=2$.}
    \label{fig:foobar}
\end{figure*}

For a fixed value of $E_{\rm p, tot}^{\rm obs}$ the photon energy output is almost independent of $\gamma_{\max}$, as indicated by the almost horizontal stripes of the same colour in Fig. \ref{fig:contourplot}. As the maximum proton Lorentz factor decreases, e.g. $\gamma_{\rm max}=10^{4}-10^{5}$, more energy has to be injected into protons to produce supercritical photon outbursts. HSC  is an efficient process for converting proton energy into photon energy, as suggested by the values of $E_{\rm \gamma, tot}^{\rm obs}/E_{\rm p, tot}^{\rm obs}$ that range between 0.03 and 0.4. Moreover, the values of the bolometric photon energy that we find fall within the range of values deduced from GRB prompt emission observations. Finally, the contours shown in the same figure indicate the ratio of the proton energy density to magnetic energy density $U_{\rm p,pk}/U_{\rm B,pk}$, at the moment of the peak of the first supercritical flare. It appears that in  the supercritical regime the condition $U_{\rm p,pk}>>U_{\rm B,pk}$ must be satisfied.

We next choose two indicative parameter sets from Fig. \ref{fig:contourplot} to look more closely into their spectra and light curves. Both cases belong to the supercritical regime and have very different $\gamma_{\max}$ and $E_{\rm p, tot}^{\rm obs}$ values (see white asterisks). We show the results for $\gamma_{\max}=10^{4.5}$ and $10^{6.5}$ on the left and right panels of  Fig.~\ref{fig:foobar}, respectively. The central plot in both panels presents snapshots of the observed broadband photon spectra during the rising part of the first pulse of the light curve as indicated in the inset plots.
Each spectrum differs from the previous one by $r_{\rm in}/c$. We use a colour scale to better illustrate the temporal evolution of the spectra in the observer's frame (see colourbar on the right of each plot). 
As it can be seen in Fig. \ref{fig:foobar}, the photon spectra show distinctive peaks coming from the reprocessing of radiation due to non linear cascades initiated by the feedback loops (see black solid line on the left panel). Therefore these spectral features are only indirectly related to the choice of $\gamma_{\max}$. For the particular example shown in the left panel, the proton luminosity required for the system to enter the supercritical regime is rather high (see Fig. \ref{fig:contourplot}), therefore the reprocessing is strong causing the photon spectrum to peak at an energy $\varepsilon_{\rm pk}^{\rm obs} $ around $\sim$MeV  in the observer's frame. For higher values of $\gamma_{\rm max}$, however, the proton luminosity requirements for entering the supercritical regime are relaxed and the photon spectral peak moves to higher energies. 

\begin{figure}
    \centering
       \includegraphics[width=\columnwidth]{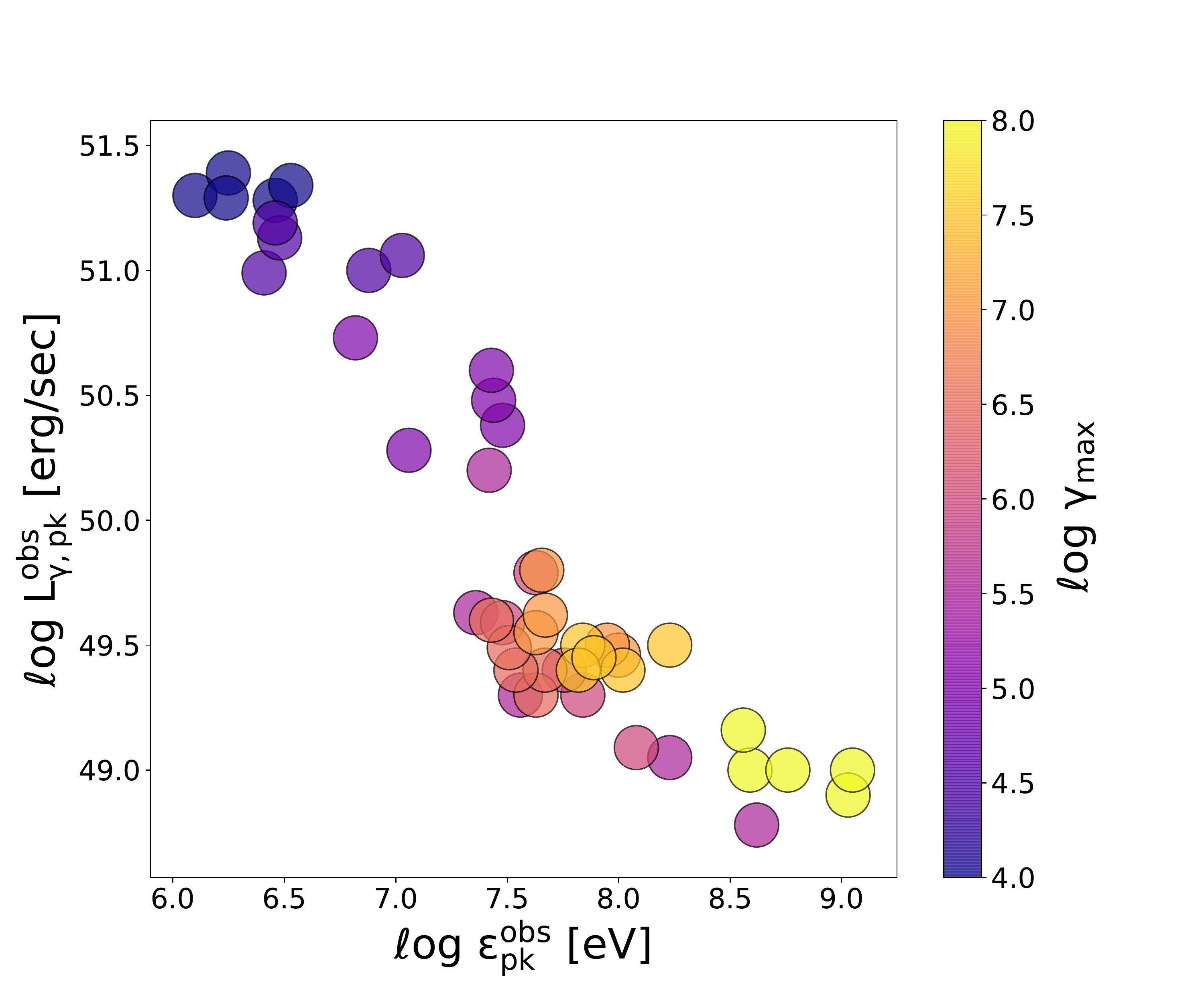}
    \caption{A plot of the photon luminosity at peak time of the first flare, $\rm L_{\gamma,pk}^{obs}$, as a function of the photon energy that corresponds to the peak of the photon spectrum released at the same moment $\rm \varepsilon_{pk}^{obs}$. All the results are computed for different values of maximum proton Lorentz factor $\gamma_{\rm max}$, shown in colour. The parameters used are the same as in Fig.~\ref{fig:foobar}.
    }
\label{fig:contourplot2}
\end{figure}

This trend becomes also evident in Fig. \ref{fig:contourplot2}, which depicts a plot of the photon luminosity at the maximum of the light curve $\rm L_{\gamma,pk}^{\rm obs}$ versus $\epsilon_{\rm pk}^{\rm obs}$, in the observer's frame, for various values of $\gamma_{\rm max}$ shown in the colour bar. Clearly the obtained photon luminosities fall in the range of the ones observed from GRBs. However, only the lower values of $\gamma_{\rm max}$ yield photon spectra with $\varepsilon_{\rm pk}^{\rm obs} \sim 1$~MeV. For $\gamma_{\rm max}\gtrsim 10^{6.5}$, the photon spectra peak at  $\varepsilon_{\rm pk}^{\rm obs} \gtrsim 10$~MeV.  For this reason, in the following section we restrict our analysis to the lower range of $\gamma_{\rm max}$ values, i.e.  $\gamma_{\rm max} \simeq 10^{4}-10^{5}$.

\subsection{The multi-blob supercritical model}
In the previous section we studied the characteristics of a photon outburst powered by HCS. We move next to apply these ideas to GRBs with light curves consisting of multiple spikes. To do so we construct synthetic light curves by allowing multiple blobs to be ejected in succession from a central source. Each one of them, depending on their initial conditions, could either enter the supercritical regime and produce one or multiple flares or could remain in the subcritical regime and show no flaring activity at all. The superposition of individual light curves produces light curves that have, as we show below, similarities to the GRB ones.

\subsubsection{The relation between the spectral peak energy and the bolometric photon energy}
Here we study how the energetics of our model compare with those of observed GRBs. We begin this analysis by assuming that a regular engine emits, every $\Delta t_{\rm off}$, $N$ expanding blobs, which have all, for the time being, the same initial set of parameters and are in the supercritical state. We also make the assumption that each blob produces a supercritical light curve that consists of a single flare. The total duration of the GRB emission is then:

\begin{equation}
    T^{\rm obs}_{\rm tot}=N(\Delta t^{\rm obs}_{\rm off}+\Delta t^{\rm obs}_{\rm 1/2})
\end{equation}
where $\Delta t^{\rm obs}_{\rm 1/2}$ is the full width at half maximum of each single flare that is determined by the model parameters, i.e. $r_{\rm in}$, $B_{\rm in}$, $u_{\rm exp}$, and $\Gamma$. We compute the emission for various parameter sets assuming a fixed $T^{\rm obs}_{\rm tot}$ (here, we set $T^{\rm obs}_{\rm tot}=60$~s). This can be achieved by changing the number of blobs or the timescale between successive blob ejections from the central engine. For each parameter set we compute the emission from a single blob and record $\Delta t^{\rm obs}_{\rm 1/2}$ and $\varepsilon_{\rm pk}^{\rm obs}$. We also  compute the bolometric photon energy $E^{\rm obs}_{\rm \gamma,tot}$ that is released during the outburst, following eq. \ref{bolenergy}. 
This is equivalent to the isotropic equivalent energy computed from the observed GRB fluence \citep[e.g.][]{Nava2008,Tu_2018}. 

GRBs are characterised by a number of correlations between different observational parameters. The Amati relation is a correlation between the GRB equivalent isotropic energy and its  rest-frame peak energy $\varepsilon_{\rm pk}^{\rm obs}$ \citep{amati}. Noting that the output parameter of our model $E^{\rm obs}_{\rm \gamma,tot}$ is equivalent to the isotropic energy entering the Amati relation we proceed with a rough comparison of the two.

\begin{figure}
    \centering
       \includegraphics[width=\columnwidth]{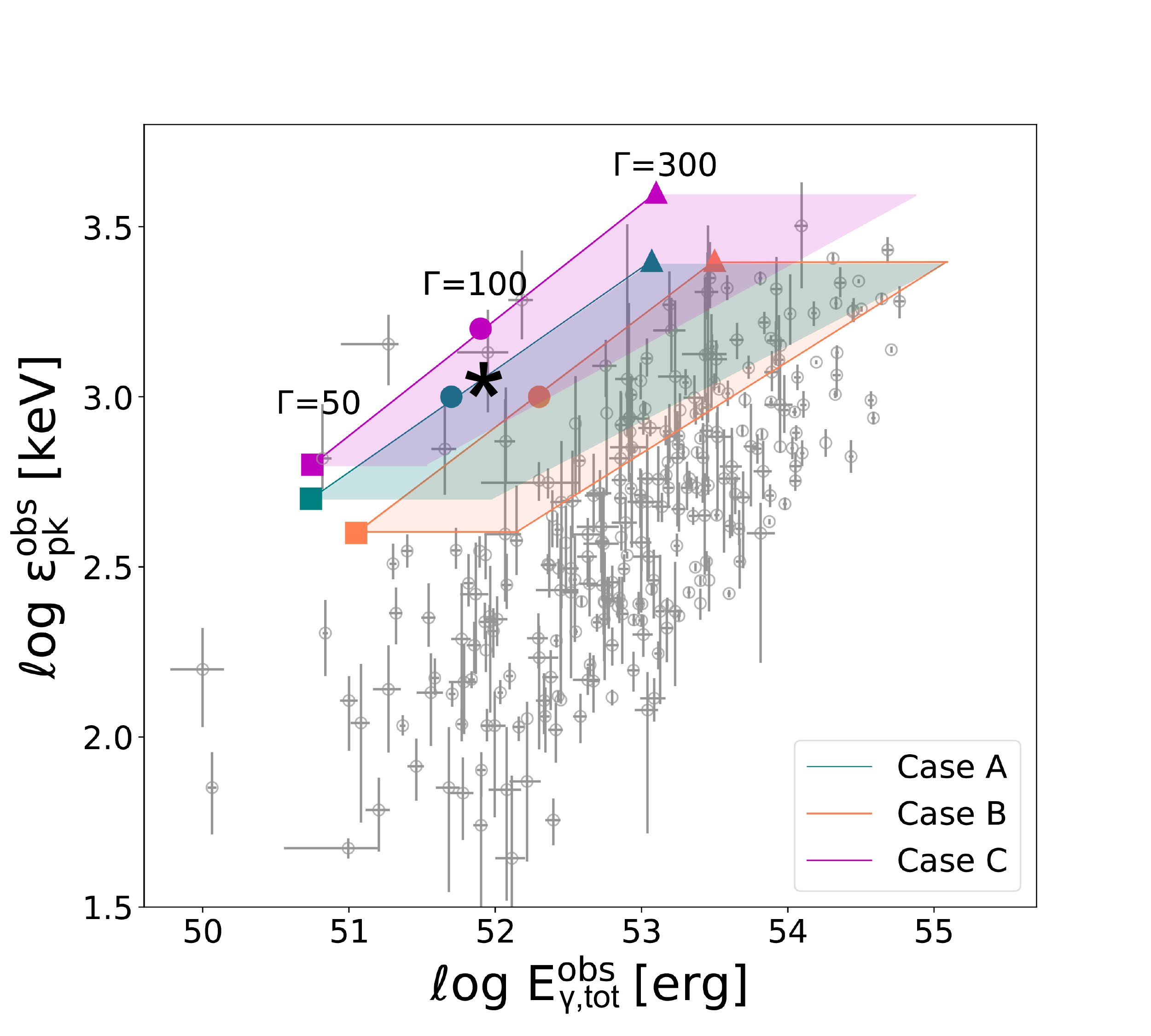}
    \caption{The Amati correlation for 275 long GRBs shown in \citet{Minaev_2019} (empty grey circles). The coloured markers correspond to the values of $E_{\rm \gamma,tot}^{\rm obs}$-$\varepsilon_{\rm pk}^{\rm obs}$ derived from our model in the case of one blob that emits a single supercritical flare. 
    We present the results for three bulk Lorentz factors, $\Gamma=300$ in triangles, $\Gamma=100$ in circles and $\Gamma=50$ in squares for three cases. In Case A, $B_{\rm in}=10^{4}$ G, $r_{\rm in}=10^{11}$ cm, $\gamma_{\rm max}=10^{4}$, $L_{\rm p,in}=10^{42.5}$ $\rm erg~ s^{-1}$, $u_{\rm exp}=10^{-2.5}c$. For the other cases, all  parameters are kept fixed except for:   $r_{\rm in}=3.16 \times 10^{11}$ cm and  $L_{\rm p,in}=10^{42.8}$ $\rm erg ~s^{-1}$ (Case B) and $u_{\rm exp}=10^{-2}c$ and  $L_{\rm p,in}=10^{42.9}$ $\rm erg~ s^{-1}$ (Case C). The light coloured regions indicate  how the results would change if multiple identical blobs were ejected from the central engine, producing a  multi-pulse light curve. The black asterisk indicates the results of the next section in the case where $\Gamma=100$.}
\label{fig:singleflare}
\end{figure}

In Fig.~\ref{fig:singleflare}  we show the values of $\varepsilon_{\rm pk}^{\rm obs}$ and $E^{\rm obs}_{\rm \gamma,tot}$ obtained by \citet{Minaev_2019} from a sample of 275 long GRBs. We overplot our results for a single blob (Case A) computed for a set of parameters from Fig.~\ref{fig:contourplot} ($B_{\rm in}=10^{4}$ G, $r_{\rm in}=10^{11}$ cm, $L_{\rm p,in}=10^{42.5}$ $\rm erg s^{-1}$, $\gamma_{\rm max}=10^{4}$, $u_{\rm exp}=10^{-2.5}c$). 
Our results depend on the choice of the bulk Lorentz factor as indicated in the figure by the different markers. For Case A specifically we find $\Delta t^{\rm obs}_{\rm 1/2}=0.8 \times (\frac{1+z}{3}) (\frac{100}{\Gamma})$~s and $\varepsilon_{\rm pk}^{\rm obs}=10^{3} \times (\frac{3}{1+z}) (\frac{\Gamma}{100})$~keV (see teal markers). The solid teal line that connects the markers corresponds to intermediate bulk Lorentz factors. The light teal coloured region indicates the increase in the isotropic photon energy in case more identical blobs were  emitted from the central engine. Assuming that $\Delta t^{\rm obs}_{\rm  1/2} \approx \Delta t^{\rm obs}_{\rm off}$, we can estimate  the maximum number of blobs which are emitted in order to observe a total light curve that lasts 60~s.  For example, $E_{\rm \gamma,tot}^{\rm obs}\simeq 2.4 \times 10^{53}$~erg would be produced, if the central engine was ejecting 37 identical blobs with $\Gamma=100$ every $0.8$~s.

We perform the same analysis by altering one of the blob parameters and compare the results. In Case B (see orange markers) we increase the initial radius of each expanding blob, i.e. $r_{\rm in}=3.16 \times 10^{11}$~cm. One needs higher proton luminosity, i.e. $L_{\rm p,in}=10^{42.9}$ erg~s$^{-1}$ in order to recapture the supercritical flare. Each blob emits as a result a higher $E^{\rm obs}_{\rm \gamma,tot}$, having however a larger $\Delta t^{\rm obs}_{\rm 1/2 }$. If $\Gamma<100$ then $\Delta t^{\rm obs}_{\rm 1/2}>4$ s, and the pulse of each blob is broad enough to create a FRED-like GRB light curve instead of a highly variable one. 

Similar results are found when the expansion velocity is altered, i.e. $u_{\rm exp}=10^{-2}c$ (see magenta markers, Case C in Fig. \ref{fig:singleflare}). In this case $\varepsilon_{\rm pk}^{\rm obs}$ is higher compared to that of Case A. This can be explained if one considers that a higher expansion velocity within the same time interval leads to a larger source, which is more optically thin, thus allowing the escape of higher energy photons.

\subsubsection{Synthetic GRB spectra and light curves}
\label{multiblob}
In order to show that HSC produces  highly variable light curves that do not depend on the randomness of the time intervals between successive blobs but on the inherent non-linearity of HSC
itself, we produce synthetic GRB light curves as follows. We assume that the blobs are ejected periodically from the central engine with slightly different initial conditions, varying in a narrow range around a set of parameters that lead to HSC.

 As an indicative example, we construct a synthetic light curve with $N=10$ blobs of initial radius $r_{\rm in}=10^{11}$~cm. These are ejected from the central engine every $50~r_{\rm in}/c$ (as measured in the comoving frame), which corresponds roughly to $\Delta t^{\rm obs}_{\rm off}=5$~s in the observer’s frame, assuming that $z=2$ and $\Gamma = 100$.

We assume that the initial magnetic field  of the source (in logarithm) is following a Gaussian distribution with a median value at $\log (B_{\rm in,med})=4.3$~G and a standard deviation $\log \sigma=0.2$. We pick randomly its value for each blob (with a resolution of 0.1 in logarithmic scale)\footnote{An even higher resolution in the selection of $\log(B)$ would not change the resulting light curves and photon spectra.}. 
 We also select values for $\log(\gamma_{\rm max})$ from a Gaussian distribution, with median value $\log(\gamma_{\rm max,med}) = 5$ and standard deviation of $\log \sigma=0.2$ . We also take into consideration that only small values of maximum proton Lorentz factors, i.e. $\gamma_{\rm max} = 10^{4}-10^{5.5}$ yield photon spectra that peak approximately at 1 MeV (see Fig. \ref{fig:contourplot2}). We calculate next the initial proton luminosity  by assuming that it is  proportional to the Poynting luminosity of the outflow, i.e. $L_{\rm p,in}=\xi B_{\rm in}^{2} r_{\rm in}^{2}c$,  where $\xi\ge 1$. Here, we choose $\xi=60$. To be consistent with the findings of the previous section,  we set  $u_{\rm exp}=10^{-2.5}c$ and inject protons in each blob until $T_{\rm B} = 400 ~ r_{\rm in}/c$; within this time at least one supercritical flare manifests  (see also Sec. \ref{sec3.3}). The model parameters are summarised in Table \ref{tab:parameters}.

\begin{figure}
    \centering
    \includegraphics[width=0.95\columnwidth]{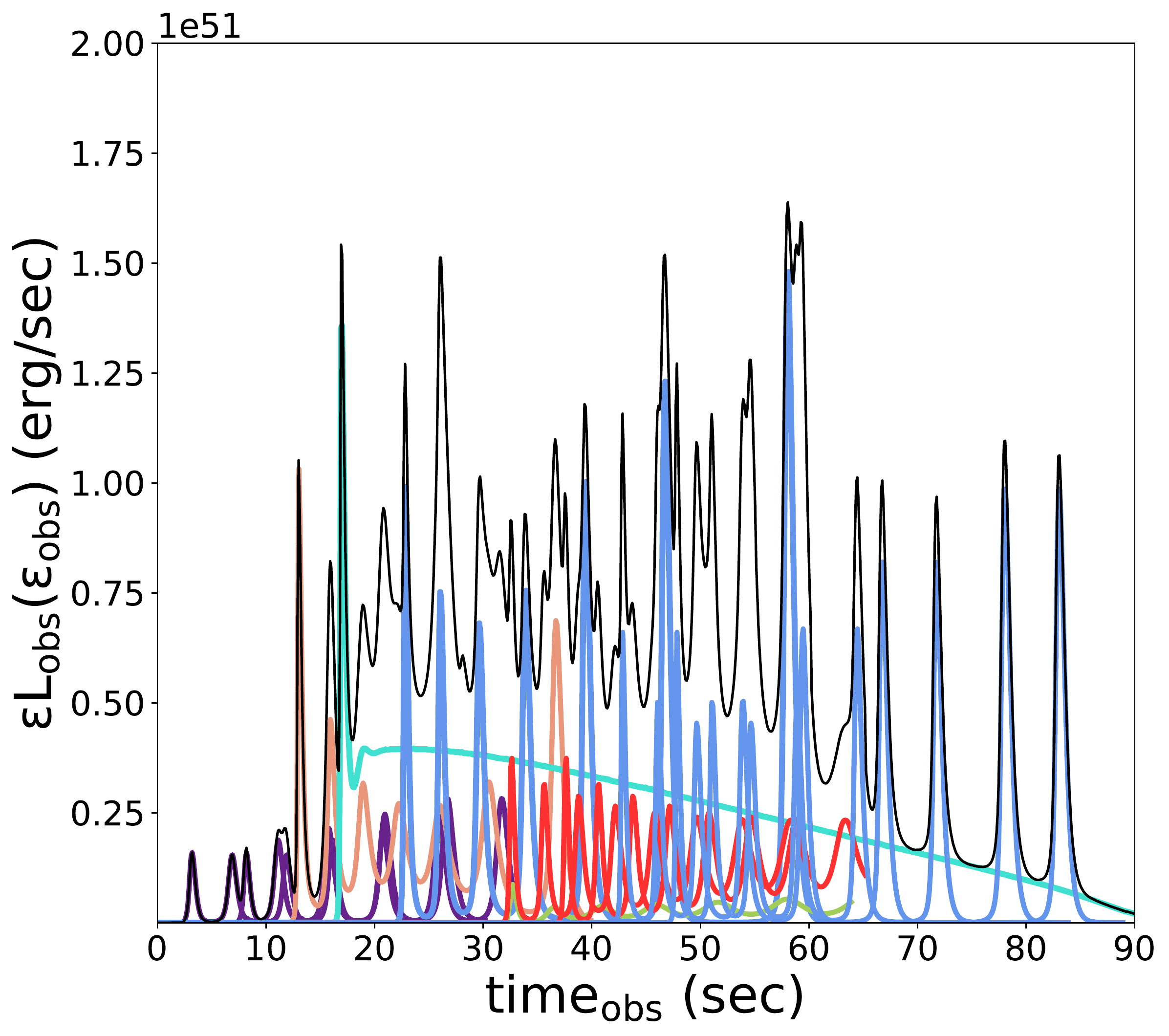}
    \includegraphics[width=\columnwidth]{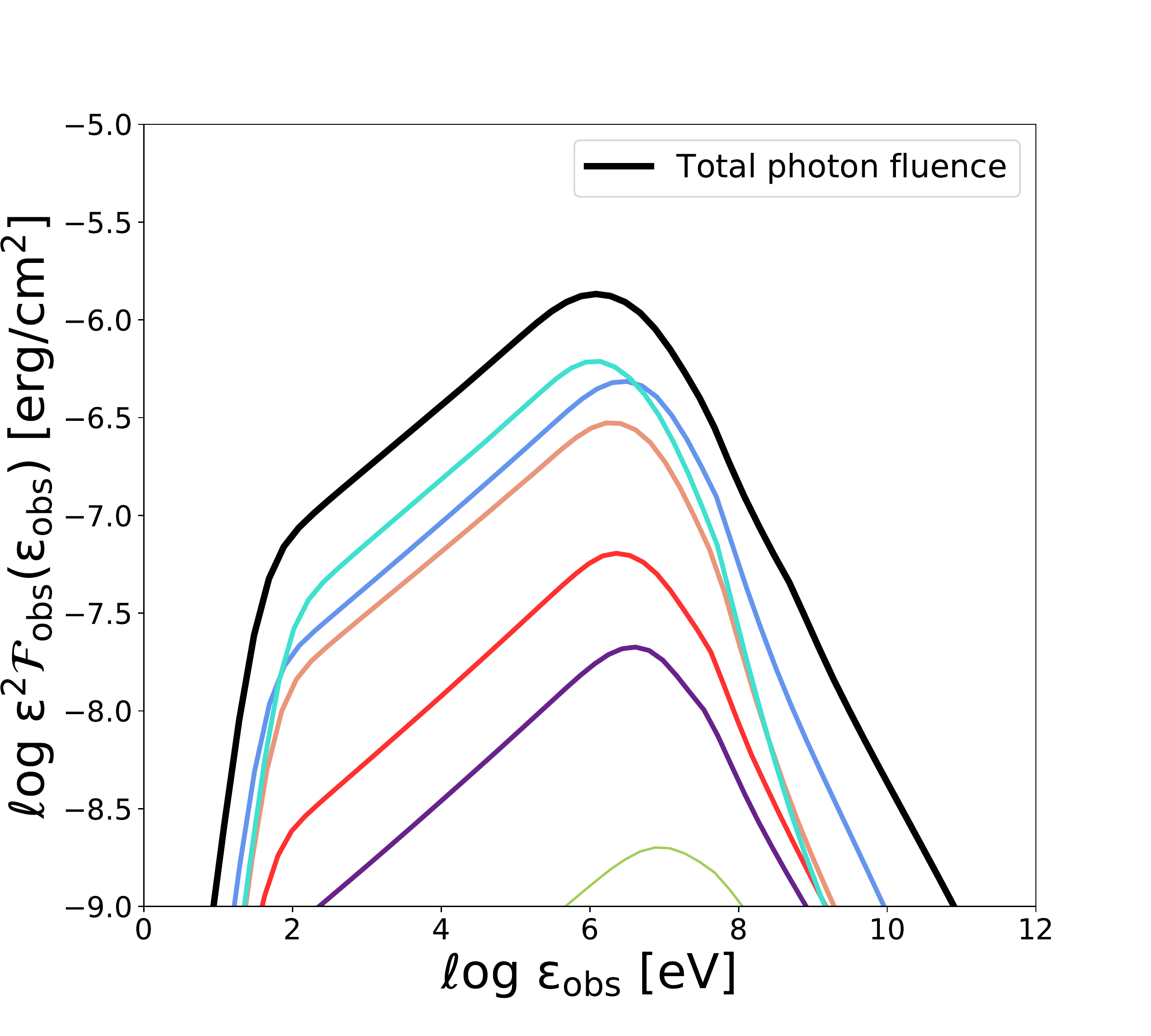}
       \caption{Upper panel: The light curve produced by the superposition of the emission of ten blobs showing supercritical behaviour (black solid line). The bulk Lorentz factor of the outflow is $\Gamma=100$. Each coloured light curve shows the correspondence of each emitting blob (see also Table \ref{tab:parameters} for the exact blob parameters). Lower panel: The total fluence per photon energy (solid black line). The coloured lines correspond to the total spectra produced by each blob and are same coloured coded as the light curves of the upper plot.}
    \label{caseA}
 \end{figure}
 \begin{table}
     \centering
      \caption{Parameters of 10 blobs used for the construction of a synthetic GRB-like light curve. Colours indicate the contribution of individual blobs to the overall light curve and spectra displayed in Figs. \ref{caseA} and \ref{neutrinospec}.}
    \begin{tabular}{c|c|c|c|r}
  \hline
  No. blobs     & $B_{\rm in}$~(G) &  $L_{\rm p,in}$~(erg s$^{-1}$) & $\gamma_{\rm max}$ & \\     \hline
  1 &$5 \times 10^{3}$ & $1.3 \times 10^{42}$ & $7.9 \times 10^{5}$ &\cellcolor{lime}  \\
    \hline
  1  &   $10^{4}$ & $2.5 \times 10^{42}$  &$1.8 \times10^{5}$ &\cellcolor{darkorchid}  \\
    \hline
 2 &   $1.6 \times 10^{4}$  & $5\times 10^{42}$&$7.9 \times 10^{4}$ &\cellcolor{indian}  \\
    \hline
 3 & $2\times10^{4}$  & $7.9 \times 10^{42}$ & $10^{5}$ &\cellcolor{cornflowerblue}  \\
  \hline
2 &  $3\times10^{4}$ & $1.25 \times 10^{43}$  &$2.2\times10^{4}$ &\cellcolor{coral}  \\
  \hline
 1 & $5\times10^{4}$&  $2.5 \times 10^{43}$ &$1.4\times10^{4}$ &\cellcolor{turquoise}  \\
  \hline
\end{tabular}

     \label{tab:parameters}
 \end{table}

We numerically compute the bolometric electromagnetic signal from each blob. The superposition of light curves from individual blobs (coloured curves) yields the synthetic light curve, which is shown with solid black line in the upper panel of Fig.~\ref{caseA}. The total flare produced is highly variable, lasts about 90 s, and the variability timescale, here defined as $\Delta t^{\rm obs}_{\rm 1/2}$,  ranges between 0.1 and 1.5~s. The radiative efficiency of the burst is $E_{\rm \gamma, tot}^{\rm obs}/E_{\rm p, tot}^{\rm obs}\approx 0.15$. 

In the lower panel of Fig. \ref{caseA} we show in colour the (differential in energy) fluence  emitted by each blob, $\mathcal{F}_{\rm obs}(\varepsilon_{\rm obs})$ $\rm [erg \ cm^{-2} \, erg^{-1}]$. We compute this by integrating the observed photon energy flux, $F_{\rm obs}(\varepsilon_{\rm obs},t_{\rm obs)}=\Gamma^3 L_{\gamma}(\varepsilon)/4 \pi d_{\rm L}^{2}$ over $T_{\rm B}$. Here, $d_{\rm L}$ is the luminosity distance of the GRB. The spectra are colour coded in the same way as the light curves of the upper panel (see also~Table~\ref{tab:parameters}). The black solid line is the superposition of the individual photon fluences. The total spectrum peaks at approximately $\varepsilon^{\rm obs}_{\rm pk}=1$~MeV, which is typical for the GRB prompt emission. The  blob with the highest value of $\gamma_{\rm max}$ (see light green curve) emits photon spectra that peak at $10$ MeV. However it does not contribute drastically to the total photon spectrum or to the total light curve because of the low value of $L_{\rm p,in}$. On the other hand, the emission from the blob with the highest value of $L_{\rm p,in}$ and lowest value of $\gamma_{\rm max}$ (see cyan curve) is imprinted on the total light curve and the peak of the photon spectrum. Furthermore, the blue light curve, which is the most variable of all, corresponds to the emission of three identical blobs. Even though their $L_{\rm p,in}$ is lower than that of the cyan curve case, the superposition of their fluences has a similar effect on the total spectrum. Overall the observed isotropic photon energy of this burst is $E^{\rm obs}_{\rm \gamma,tot}\simeq 10^{52}$~erg ($2.5\times10^{53}$~erg) for $\Gamma=100$ ($300$), which places it in the upper part of the $\varepsilon_{\rm pk}^{\rm obs}-E_{\rm \gamma,tot}^{\rm obs}$ diagram (see the black asterisk in Fig.~\ref{fig:singleflare} for the case of $\Gamma=100$ ).
The results are compatible with a typical long GRB.

\subsubsection{The total neutrino fluence}

\begin{figure}
    \centering
    \includegraphics[width=1.01\columnwidth]{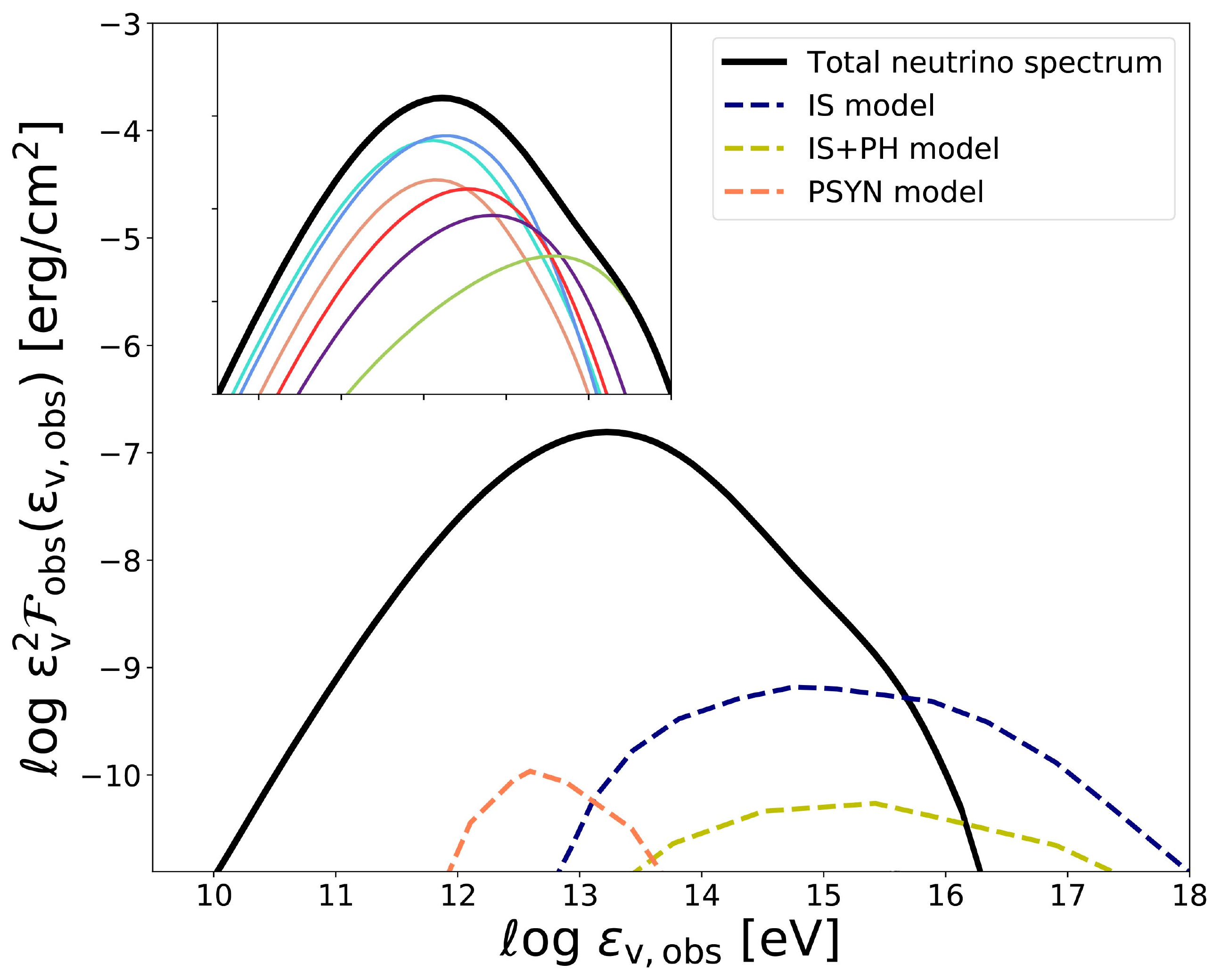}
    \caption{The all-flavour neutrino fluence (solid black line) calculated for a HSC burst with $10$ expanding blobs. In the inset plot we show the neutrino fluence observed from each expanding blob, same colour coded as the photon spectra and light curves of Fig.~\ref{caseA}. The dashed lines correspond to neutrino spectra computed with the internal shock (IS), the internal shock+photospheric dissipation (IS+PH) and the proton synchrotron (PSYN) models, as shown in \citet{pitik21}.}
    \label{neutrinospec}
\end{figure}

High-energy neutrinos are a guaranteed by-product of the HSC model~\citep[see also][]{14MPGM}. It is therefore interesting to compare the neutrino predictions of our model with other existing ones. In order to compute the all-flavour neutrino spectra emitted by each expanding blob, we utilise the \cite{DM12} numerical code that treats in detail neutrino emissivities using the event generator SOPHIA~\citep{2000muck}.  
Coloured curves in Fig. \ref{neutrinospec} show the contribution from individual blobs to the total neutrino fluence (black line). Here, the same colour coding as in Fig. \ref{caseA} is used. The total neutrino emission of the burst peaks at $\simeq 10$~TeV and is dominated mostly by the cyan and blue spectra. The cyan spectrum, in particular, which corresponds to the blob with the lowest $\gamma_{\rm max}$-highest $L_{\rm p,in}$ values, controls the total peak energy and fluence. On the other hand the green spectrum, which correspond to the blob with the highest $\gamma_{\rm max}$-lowest $L_{\rm p,in}$ values, contributes mostly to the highest part of the neutrino spectrum energies. The ratio of the peaks of the total photon to all-flavour neutrino fluences in this example case is approximately $\mathcal{F}^{\rm obs}_{\rm \gamma ,pk}/\mathcal{F}^{\rm obs}_{\rm \nu,pk}\approx12.5 $. 

In Fig. \ref{neutrinospec} we also show the predicted neutrino spectra shown in \cite{pitik21}, who computed the neutrino production for a benchmark high-luminosity GRB, among others in the internal shock model  \citep[IS,][]{Rees_1994}, in the internal shock model with dissipative photosphere \citep[IS+PH,][]{PHIS} and in the proton synchrotron marginally fast cooling model  \citep[PSYN,][]{florou2021marginally, pitik21}. Because these models were computed for a GRB with a different $\gamma$-ray fluence than ours, we appropriately rescaled the neutrino fluence spectra of \cite{pitik21} for a fair comparison to our results. These models are overplotted with  dashed dark blue, yellow and orange curves respectively. The neutrino spectrum of the HSC model is about two orders of magnitude higher than the expected neutrino spectrum of the IS model, and peaks at about two orders of magnitude lower energy. The lower peak energy and higher fluence found in the HSC model compared to the IS scenarios are a direct consequence of the fact that the HSC model for GRBs requires relatively low  energies of $\gamma_{\rm max}$  (see Sec. \ref{Sec:3}) and is characterised by high neutrino production efficiency~\citep[see e.g.][]{14MPGM}. The low values of $\gamma_{\max}$ that are needed to explain the MeV photon spectrum with proton synchrotron radiation in the PSYN scenario lead to similar peak neutrino energies to the HSC model. Meanwhile, the neutrino production efficiency in the PSYN scenario is much lower than in the HSC model, since protons are cooling via synchrotron radiation in the presence of very strong magnetic fields in the former case.

\begin{table*}
\centering
\caption{A summary of the  results in this section which are compatible with the GRB phenomenology. The symbol $\uparrow$ indicates increase, while the symbol $\downarrow$ indicates decrease. Furthermore we sign with {\color{teal} \checkmark } the parameters that agree with the GRB phenomenology and with {\color{teal} \xmark } those that do not agree. We denote the multiple bursts with MB and the single burst with SB. }
\label{tablesum}
\begin{tabular}{ |p{2cm}||p{2.2cm}|p{3.3cm}|p{3.8cm}||p{2.5cm}| }
 \hline
 \multicolumn{5}{|c|}{Results}  \\
 \hline
 Parameters& Modification &Light curves (LC)& Photon Spectra (SED) & Relevance to GRBs\\
 \hline

$u_{\rm exp}/c$ &   $u_{\rm exp} \uparrow$ &  MB $\xrightarrow[]{}$ SB , &$\varepsilon^{\rm obs}_{\rm pk}$ $\uparrow$ & LC  {\color{teal} \checkmark }, SED {\color{teal} \xmark } \\
$B \sim r^{-q}$ &  $q \uparrow$ & MB $\downarrow$ & $\varepsilon^{\rm obs}_{\rm pk}\approx$ constant & LC {\color{teal} \checkmark }, SED {\color{teal} \checkmark }\\
$\gamma_{\rm max}$& $\gamma_{\rm max} \uparrow$   & MB $\xrightarrow[]{}$ SB $\xrightarrow[]{}$ MB  &  $\varepsilon^{\rm obs}_{\rm pk}\uparrow$   & LC  {\color{teal} \checkmark }, SED {\color{teal} \xmark } \\
 \hline
\end{tabular}
\end{table*}

\section{Summary \& Discussion}
\label{sec5}
In this paper we have extended recent work on HSC by examining its {\sl modus operandi} in expanding sources. Prompted by our results we have applied them to the GRB prompt emission. We summarize next our basic results and discuss some aspects of our model.

HSC  is a universal property of relativistic proton plasmas and appears {\sl spontaneously} once appropriate conditions are satisfied. Its physical premise is based on a gradual proton accumulation inside the source and a sudden dissipation of the stored energy via one or more photon outbursts, bearing similarities to nuclear piles. These outbursts are, therefore, a {\sl natural} consequence of the phase transition that the hadronic plasma undergoes and appear even when all parameters are kept constant. 

HSC can occur only for specific combinations of initial radii, magnetic fields and proton energies. Other parameters like the power-law indices of the magnetic field and proton luminosity radial profiles as well as the power-law slope of the proton energy distribution play only an auxiliary role, as they do not qualitatively change the general picture described above. Determination of the critical conditions leading to HSC is still not sufficient for fully understanding the problem under study.  The non-linear stages of HSC require a fully numerical treatment of the physical processes coupling photons with relativistic protons. Due to the complexity of the problem, there are only empirical ways of mapping the initial critical conditions to the final outcome. These involve a plethora of temporal and spectral signatures that can be tested readily against observations.

The role of adiabatic expansion to the manifestation of HSC has not been explored so far. We therefore studied  thoroughly the effects of the adiabatic expansion velocity on the phenomenology of HSC. Small expansion velocities favour short-duration spiky outbursts characterised by high radiative efficiency. As the expansion velocity becomes larger, the bursts appear longer, smoother and less efficient. Also the multiplicity of bursts decreases with increasing expansion velocity,  degenerating into a single burst  when $u_{\rm exp} \gtrsim 0.01c$. Single bursts can still be produced up to very high expansion velocities $\sim 0.5c$ (see e.g. Fig.~\ref{fig:2}). In other words, HSC persists for all expansion velocities, a fact that is impressive by itself if one considers that its onset depends on the proton column density. It is exactly because of this dependence, that higher initial proton luminosities are needed as $u_{\rm exp}$ becomes larger. \cite{am20} showed that the energy density in relativistic protons $U_{\rm p}$ greatly exceeds the energy density in magnetic fields $U_{\rm B}$ for stationary sources in HSC. This result  also holds for expanding sources where both energy densities vary continuously. Systems where $U_{\rm B}\simeq U_{\rm p}$ never become supercritical. Furthermore, although the choice of a steeper magnetic field profile, e.g. $1/r^{2}$, for fixed initial proton luminosities  and expansion velocities, makes the system more luminosity demanding, overall it leads to similar results as the flatter case of $1/r$.

The above epitomise some of the key results of HSC in expanding sources in the absence of external radiation fields and injection of relativistic (primary) electrons in the source. Both aspects have been shown to have a stabilising effect on the non-linear development of HSC in stationary sources \citep{am20}, which also apply to expanding systems. For instance, the system reaches quickly (within a few source light-crossing times) a steady state characterised by high radiative efficiency without exhibiting bursty behaviour, if the injection luminosity of primary electrons exceeds $\sim 10\%$ of the proton luminosity. Similarly, if the density of external radiation fields is higher than the intrinsic photon density, then the non-linear coupling between protons and their own radiation is weakened. In light of these results we continue our discussion on the relevance of HSC to the GRB prompt emission. 

We find that emitting regions (blobs) with initial radii $\sim 10^{11}-10^{12}~\rm cm$ and maximum proton Lorentz factors  $\sim 10^4-10^5$ produce outbursts with characteristics that are similar to those observed from GRBs during the prompt phase. For small expansion velocities the light curves consist of multiple spikes with typical duration of a few tenths of the second and spectra that peak at $\sim$MeV energies. Both features are consistent with the GRB phenomenology. As explained earlier, there is no {\sl a-priori} reason that this should be the case. Interestingly, the expansion velocity also controls the appearance of the light curve, as higher expansion velocities give rise to single-pulse outbursts that resemble FRED-like GRB light curves.  Because the photon pulses from a single slowly expanding blob are separated by a few hundreds source light-crossing times, one needs to superimpose the emission of a few blobs to obtain multi-pulse light curves, as those observed in many GRBs. Even small variations in the initial conditions between successive blobs can produce highly variable light curves (see Table~\ref{tab:parameters} and Fig.~\ref{caseA}). 

The expansion velocity also affects the peak energy of the photon spectrum, $\varepsilon_{\rm pk}^{\rm obs}$. Faster expansion of the source roughly corresponds to a lower intrinsic opacity to $\gamma \gamma$ pair production, thus pushing $\varepsilon_{\rm pk}^{\rm obs}$ beyond the MeV range for all other parameters kept fixed (see Fig.~\ref{fig:singleflare}). Another way to obtain photon spectra peaking in the GeV range with our model is to consider the injection of protons with higher maximum energies (see Figs.~\ref{fig:foobar}). The prediction of the HSC model is that if such GRBs do exist, they should be on average less luminous than their MeV counterparts, as illustrated in Fig.~\ref{fig:contourplot2}. The maximum proton energy also affects the appearance of the light curve, with multi-pulse light curves produced from a single blob containing protons with $\gamma_{\max}\simeq 10^{4}-10^{5}$ or $\gtrsim 10^{6.5}$. For intermediate values, the outbursts usually consist of a single pulse. Nonetheless, the HSC model cannot explain GRBs with soft spectra peaking in the 10--100 keV range.

Some of the above results are summarised in Table \ref{tablesum}, where we show the dependence of the supercritical temporal and spectral behaviour on the expansion velocity, the magnetic field profile and the maximum Lorentz factor. We note with $\uparrow$ ($\downarrow$) the parameter increase (decrease). We also compare our results with the GRB phenomenology, denoting with {\color{teal} \checkmark } the cases were the light curves and the photon spectra bear broad similarities with the observations. 
 
 Since the photon spectrum at the peak time of the outburst is a result of intense electromagnetic cascades from secondary pairs and $\gamma$-ray photons, we expect that TeV photons will be severely attenuated inside the source and the ratios of spectral luminosities between TeV and MeV can be as low as $\sim 10^{-5}$ (see e.g. Fig.~\ref{fig:foobar}). TeV photons will also be attenuated en route to us by the photons of the extragalactic background light (EBL). Therefore, the HSC model suggests that detection of prompt GRB emission at TeV energies would be challenging even for sensitive instruments such as the Cherenkov Telescope Array \citep[CTA,][]{cta}. For example, the average photon spectrum of the indicative burst shown in Fig.~\ref{caseA} (lower panel) exceeds the 50 hr CTA sensitivity curve at 1 TeV only for $z\lesssim 0.3$ (assuming the EBL model of \citet{2010finke}).

Besides electromagnetic radiation high energy neutrinos are copiously produced in HSC outbursts. For those in particular that are powered by protons of relatively low energies (e.g. $\gamma_{\max}=10^4-10^5$), we find that the all-flavour neutrino spectrum peaks at $\sim 10 \, {\rm TeV} \, (3/(1+z))$ and has a peak fluence of about $10\%$ of the peak photon energy fluence (see lower panel in Fig.~\ref{caseA} and \ref{neutrinospec}). The neutrino-to-photon peak fluence ratio can range between 3\% and 30\% depending on the main model parameters, like $u_{\rm exp}, \gamma_{\max}$ and $B$. This ratio however cannot be fully determined {\sl a priori} by simply selecting the appropriate initial conditions, as it is the outcome of the proton-photon interactions in the non-linear stages of HSC. As a result, the neutrino predictions of the HSC model for GRBs are substantially different than those typically found in the literature \citep{Gao_2012,BAERWALD201566,https://doi.org/10.48550/arxiv.1511.01396,Bustamante_2017,Biehl_2018,pitik21,florou2021marginally}.

We can also compare the predictions of the HSC model with the stacking flux limits from the IceCube neutrino telescope. To estimate the all-sky quasi-diffuse flux of muon neutrinos and antineutrinos we assume that our benchmark long-duration GRB at $z=2$ with $E^{\rm obs}_{\rm \gamma, tot}\simeq 10^{52}$~erg s$^{-1}$ and $\Gamma=100$ (see Sec.~\ref{multiblob}) is representative of the entire GRB population. Given a rate of long GRBs $\dot{N} = 667$~yr$^{-1}$ \citep{2017ApJ...843..112A}, the stacking flux for muon neutrinos over the whole sky can be written as $\Phi_{\nu_\mu+\bar{\nu}_\mu} \simeq (\dot{N}/4 \pi) (\mathcal{F}_{\nu +\bar{\nu}}/3)$, where we assumed vacuum flavour mixing. Using $\mathcal{F}_{\nu +\bar{\nu}} \simeq 10^{-7}$~erg~cm$^{-2}$ as a conservative value (i.e. 10\% of the $\gamma$-ray fluence), we find $\Phi_{\nu_\mu+\bar{\nu}_\mu} \simeq 3.7\times10^{-11}$~GeV cm$^{-2}$ s$^{-1}$ sr$^{-1}$. With a peak energy around 10~TeV our prediction is well below the  IceCube stacking limit \citep{2017ApJ...843..112A} in agreement with the non-detection of high-energy neutrinos from targeted GRB searches. It is however very close to the projected limit for IceCube-Gen2 in the energy range of 10--100~TeV \citep{IceCube-Gen2}, suggesting that the HSC scenario could be testable within the next decade with neutrino observations.

\section*{Acknowledgements}
IF acknowledges that this research is co-financed by Greece and the European Union (European Social Fund- ESF) through the Operational
Programme «Human Resources Development, Education and
Lifelong Learning» in the context of the project
“Strengthening Human Resources Research Potential via
Doctorate Research – $2^{\rm nd}$ Cycle” (MIS-5000432), implemented by the State Scholarships Foundation (IKY). MP and IF acknowledge support from the MERAC Foundation through the project THRILL.

\section*{Data Availability}
The data shown in Fig.~\ref{fig:singleflare} are adopted from \cite{Minaev_2019}. All numerical models presented in this paper were computed using a proprietary numerical code. They can be shared upon reasonable request to the authors.

\bibliographystyle{mnras}
\bibliography{bibliography} 

\begin{thebibliography}{}
\makeatletter
\relax
\def\mn@urlcharsother{\let\do\@makeother \do\$\do\&\do\#\do\^\do\_\do\%\do\~}
\def\mn@doi{\begingroup\mn@urlcharsother \@ifnextchar [ {\mn@doi@}
  {\mn@doi@[]}}
\def\mn@doi@[#1]#2{\def\@tempa{#1}\ifx\@tempa\@empty \href
  {http://dx.doi.org/#2} {doi:#2}\else \href {http://dx.doi.org/#2} {#1}\fi
  \endgroup}
\def\mn@eprint#1#2{\mn@eprint@#1:#2::\@nil}
\def\mn@eprint@arXiv#1{\href {http://arxiv.org/abs/#1} {{\tt arXiv:#1}}}
\def\mn@eprint@dblp#1{\href {http://dblp.uni-trier.de/rec/bibtex/#1.xml}
  {dblp:#1}}
\def\mn@eprint@#1:#2:#3:#4\@nil{\def\@tempa {#1}\def\@tempb {#2}\def\@tempc
  {#3}\ifx \@tempc \@empty \let \@tempc \@tempb \let \@tempb \@tempa \fi \ifx
  \@tempb \@empty \def\@tempb {arXiv}\fi \@ifundefined
  {mn@eprint@\@tempb}{\@tempb:\@tempc}{\expandafter \expandafter \csname
  mn@eprint@\@tempb\endcsname \expandafter{\@tempc}}}

\bibitem[\protect\citeauthoryear{{Aartsen} et~al.,}{{Aartsen}
  et~al.}{2017}]{2017ApJ...843..112A}
{Aartsen} M.~G.,  et~al., 2017, \mn@doi [\apj] {10.3847/1538-4357/aa7569},
  \href {https://ui.adsabs.harvard.edu/abs/2017ApJ...843..112A} {843, 112}

\bibitem[\protect\citeauthoryear{Aartsen et~al.,}{Aartsen
  et~al.}{2021a}]{Aartsen_2021}
Aartsen M.~G.,  et~al., 2021a, \mn@doi [Journal of Physics G: Nuclear and
  Particle Physics] {10.1088/1361-6471/abbd48}, 48, 060501

\bibitem[\protect\citeauthoryear{{Aartsen} et~al.,}{{Aartsen}
  et~al.}{2021b}]{IceCube-Gen2}
{Aartsen} M.~G.,  et~al., 2021b, \mn@doi [Journal of Physics G Nuclear Physics]
  {10.1088/1361-6471/abbd48}, \href
  {https://ui.adsabs.harvard.edu/abs/2021JPhG...48f0501A} {48, 060501}

\bibitem[\protect\citeauthoryear{Abdalla et~al.,}{Abdalla
  et~al.}{2019}]{Abdalla_2019}
Abdalla H.,  et~al., 2019, \mn@doi [Nature] {10.1038/s41586-019-1743-9}, 575,
  464–467

\bibitem[\protect\citeauthoryear{{Abdo} et~al.,}{{Abdo}
  et~al.}{2009}]{2009Abdo}
{Abdo} A.~A.,  et~al., 2009, \mn@doi [\apjl] {10.1088/0004-637X/706/1/L138},
  \href {https://ui.adsabs.harvard.edu/abs/2009ApJ...706L.138A} {706, L138}

\bibitem[\protect\citeauthoryear{Ackermann et~al.,}{Ackermann
  et~al.}{2011}]{Ackermann_2011}
Ackermann M.,  et~al., 2011, \mn@doi [The Astrophysical Journal]
  {10.1088/0004-637x/729/2/114}, 729, 114

\bibitem[\protect\citeauthoryear{{Amati, L.} et~al.,}{{Amati, L.}
  et~al.}{2002}]{amati}
{Amati, L.} et~al., 2002, \mn@doi [A\&A] {10.1051/0004-6361:20020722}, 390, 81

\bibitem[\protect\citeauthoryear{{Asano} \& {Inoue}}{{Asano} \&
  {Inoue}}{2007}]{Asano2007}
{Asano} K.,  {Inoue} S.,  2007, \mn@doi [\apj] {10.1086/522939}, \href
  {https://ui.adsabs.harvard.edu/abs/2007ApJ...671..645A} {671, 645}

\bibitem[\protect\citeauthoryear{Asano, Guiriec  \& M{\'{e}}sz{\'{a}}ros}{Asano
  et~al.}{2009}]{Asano_2009}
Asano K.,  Guiriec S.,   M{\'{e}}sz{\'{a}}ros P.,  2009, \mn@doi [The
  Astrophysical Journal] {10.1088/0004-637x/705/2/l191}, 705, L191

\bibitem[\protect\citeauthoryear{Baerwald, Bustamante  \& Winter}{Baerwald
  et~al.}{2015}]{BAERWALD201566}
Baerwald P.,  Bustamante M.,   Winter W.,  2015, \mn@doi [Astroparticle
  Physics] {https://doi.org/10.1016/j.astropartphys.2014.07.007}, 62, 66

\bibitem[\protect\citeauthoryear{{Band} et~al.,}{{Band} et~al.}{1993}]{Band}
{Band} D.,  et~al., 1993, \mn@doi [\apj] {10.1086/172995}, \href
  {https://ui.adsabs.harvard.edu/abs/1993ApJ...413..281B} {413, 281}

\bibitem[\protect\citeauthoryear{Beloborodov \& Mészáros}{Beloborodov \&
  Mészáros}{2017}]{Beloborodov_2017}
Beloborodov A.~M.,  Mészáros P.,  2017, \mn@doi [Space Science Reviews]
  {10.1007/s11214-017-0348-6}, 207, 87–110

\bibitem[\protect\citeauthoryear{Biehl, Boncioli, Lunardini  \& Winter}{Biehl
  et~al.}{2018}]{Biehl_2018}
Biehl D.,  Boncioli D.,  Lunardini C.,   Winter W.,  2018, \mn@doi [Scientific
  Reports] {10.1038/s41598-018-29022-4}, 8

\bibitem[\protect\citeauthoryear{{Boula} \& {Mastichiadis}}{{Boula} \&
  {Mastichiadis}}{2022}]{BoulaMasti}
{Boula} S.,  {Mastichiadis} A.,  2022, \mn@doi [\aap]
  {10.1051/0004-6361/202142126}, \href
  {https://ui.adsabs.harvard.edu/abs/2022A&A...657A..20B} {657, A20}

\bibitem[\protect\citeauthoryear{{Burgess}, {B{\'e}gu{\'e}}, {Greiner},
  {Giannios}, {Bacelj}  \& {Berlato}}{{Burgess} et~al.}{2020}]{2020burges}
{Burgess} J.~M.,  {B{\'e}gu{\'e}} D.,  {Greiner} J.,  {Giannios} D.,  {Bacelj}
  A.,   {Berlato} F.,  2020, \mn@doi [Nature Astronomy]
  {10.1038/s41550-019-0911-z}, \href
  {https://ui.adsabs.harvard.edu/abs/2020NatAs...4..174B} {4, 174}

\bibitem[\protect\citeauthoryear{Bustamante, Heinze, Murase  \&
  Winter}{Bustamante et~al.}{2017}]{Bustamante_2017}
Bustamante M.,  Heinze J.,  Murase K.,   Winter W.,  2017, \mn@doi [The
  Astrophysical Journal] {10.3847/1538-4357/837/1/33}, 837, 33

\bibitem[\protect\citeauthoryear{Crider et~al.,}{Crider
  et~al.}{1997}]{Crider_1997}
Crider A.,  et~al., 1997, \mn@doi [The Astrophysical Journal] {10.1086/310574},
  479, L39

\bibitem[\protect\citeauthoryear{Dermer \& Atoyan}{Dermer \&
  Atoyan}{2003}]{dermer-attonian2003}
Dermer C.~D.,  Atoyan A.,  2003, \mn@doi [Phys. Rev. Lett.]
  {10.1103/PhysRevLett.91.071102}, 91, 071102

\bibitem[\protect\citeauthoryear{Dermer \& Atoyan}{Dermer \&
  Atoyan}{2006}]{Dermer-atonian_2006}
Dermer C.~D.,  Atoyan A.,  2006, \mn@doi [New Journal of Physics]
  {10.1088/1367-2630/8/7/122}, 8, 122

\bibitem[\protect\citeauthoryear{{Dimitrakoudis}, {Mastichiadis, A.},
  {Protheroe, R. J.}  \& {Reimer, A.}}{{Dimitrakoudis} et~al.}{2012}]{DM12}
{Dimitrakoudis} {Mastichiadis, A.} {Protheroe, R. J.}  {Reimer, A.} 2012,
  \mn@doi [A\&A] {10.1051/0004-6361/201219770}, 546, A120

\bibitem[\protect\citeauthoryear{Finke, Razzaque  \& Dermer}{Finke
  et~al.}{2010}]{2010finke}
Finke J.~D.,  Razzaque S.,   Dermer C.~D.,  2010, \mn@doi [The Astrophysical
  Journal] {10.1088/0004-637x/712/1/238}, 712, 238–249

\bibitem[\protect\citeauthoryear{Florou, Petropoulou  \& Mastichiadis}{Florou
  et~al.}{2021}]{florou2021marginally}
Florou I.,  Petropoulou M.,   Mastichiadis A.,  2021, \mn@doi [Monthly Notices
  of the Royal Astronomical Society] {10.1093/mnras/stab1285}, 505, 1367

\bibitem[\protect\citeauthoryear{Gao, Asano  \& M{\'{e}}sz{\'{a}}ros}{Gao
  et~al.}{2012}]{Gao_2012}
Gao S.,  Asano K.,   M{\'{e}}sz{\'{a}}ros P.,  2012, \mn@doi [Journal of
  Cosmology and Astroparticle Physics] {10.1088/1475-7516/2012/11/058}, 2012,
  058

\bibitem[\protect\citeauthoryear{Gendre et~al.,}{Gendre
  et~al.}{2013}]{Gendre_2013}
Gendre B.,  et~al., 2013, \mn@doi [The Astrophysical Journal]
  {10.1088/0004-637x/766/1/30}, 766, 30

\bibitem[\protect\citeauthoryear{{Ghisellini} et~al.,}{{Ghisellini}
  et~al.}{2020}]{ghisselini20}
{Ghisellini} et~al., 2020, \mn@doi [A\&A] {10.1051/0004-6361/201937244}, 636,
  A82

\bibitem[\protect\citeauthoryear{Goldstein et~al.,}{Goldstein
  et~al.}{2012}]{Goldstein_2012}
Goldstein A.,  et~al., 2012, \mn@doi [The Astrophysical Journal Supplement
  Series] {10.1088/0067-0049/199/1/19}, 199, 19

\bibitem[\protect\citeauthoryear{{Guiriec}}{{Guiriec}}{2012}]{2012Guiriec}
{Guiriec} S.,  2012, in 39th COSPAR Scientific Assembly. p.~682

\bibitem[\protect\citeauthoryear{Hinshaw et~al.,}{Hinshaw et~al.}{2013}]{wmap9}
Hinshaw G.,  et~al., 2013, \mn@doi [The Astrophysical Journal Supplement
  Series] {10.1088/0067-0049/208/2/19}, 208, 19

\bibitem[\protect\citeauthoryear{{Kardashev}}{{Kardashev}}{1962}]{kardashev}
{Kardashev} N.~S.,  1962, \azh, \href
  {https://ui.adsabs.harvard.edu/abs/1962AZh....39..393K} {39, 393}

\bibitem[\protect\citeauthoryear{{Katz}}{{Katz}}{1994}]{1994katz}
{Katz} J.~I.,  1994, \mn@doi [\apjl] {10.1086/187523}, \href
  {https://ui.adsabs.harvard.edu/abs/1994ApJ...432L.107K} {432, L107}

\bibitem[\protect\citeauthoryear{{Kazanas}, {Georganopoulos}  \&
  {Mastichiadis}}{{Kazanas} et~al.}{2002}]{Kazanas02}
{Kazanas} D.,  {Georganopoulos} M.,   {Mastichiadis} A.,  2002, \mn@doi [\apjl]
  {10.1086/344518}, \href {http://adsabs.harvard.edu/abs/2002ApJ...578L..15K}
  {578, L15}

\bibitem[\protect\citeauthoryear{{Kirk} \& {Mastichiadis}}{{Kirk} \&
  {Mastichiadis}}{1992}]{KM92}
{Kirk} J.~G.,  {Mastichiadis} A.,  1992, \mn@doi [\nat] {10.1038/360135a0},
  \href {http://adsabs.harvard.edu/abs/1992Natur.360..135K} {360, 135}

\bibitem[\protect\citeauthoryear{Knödlseder}{Knödlseder}{2020}]{cta}
Knödlseder J.,  2020, The Cherenkov Telescope Array (\mn@eprint {arXiv}
  {2004.09213})

\bibitem[\protect\citeauthoryear{Kumar \& Zhang}{Kumar \&
  Zhang}{2015}]{KUMAR2015}
Kumar P.,  Zhang B.,  2015, \mn@doi [Physics Reports]
  {https://doi.org/10.1016/j.physrep.2014.09.008}, 561, 1

\bibitem[\protect\citeauthoryear{{MAGIC Collaboration} et~al.,}{{MAGIC
  Collaboration} et~al.}{2019}]{2019magiccoll}
{MAGIC Collaboration} et~al., 2019, \mn@doi [\nat] {10.1038/s41586-019-1750-x},
  \href {https://ui.adsabs.harvard.edu/abs/2019Natur.575..455M} {575, 455}

\bibitem[\protect\citeauthoryear{{Mastichiadis} \& {Kazanas}}{{Mastichiadis} \&
  {Kazanas}}{2009}]{KaM}
{Mastichiadis} A.,  {Kazanas} D.,  Mar 2009, \apjl

\bibitem[\protect\citeauthoryear{{Mastichiadis} \& {Kirk}}{{Mastichiadis} \&
  {Kirk}}{1995}]{95km}
{Mastichiadis} A.,  {Kirk} J.~G.,  1995, \aap, \href
  {https://ui.adsabs.harvard.edu/abs/1995A%26A...295..613M} {295, 613}

\bibitem[\protect\citeauthoryear{{Mastichiadis} \& {Kirk}}{{Mastichiadis} \&
  {Kirk}}{1997}]{97km}
{Mastichiadis} A.,  {Kirk} J.~G.,  1997, \aap, \href
  {https://ui.adsabs.harvard.edu/abs/1997A%26A...320...19M} {320, 19}

\bibitem[\protect\citeauthoryear{Mastichiadis, Florou, Kefala, Boula  \&
  Petropoulou}{Mastichiadis et~al.}{2020}]{am20}
Mastichiadis A.,  Florou I.,  Kefala E.,  Boula S.~S.,   Petropoulou M.,  2020,
  \mn@doi [Monthly Notices of the Royal Astronomical Society]
  {10.1093/mnras/staa1308}, 495, 2458–2474

\bibitem[\protect\citeauthoryear{Meszaros, Rees  \& Papathanassiou}{Meszaros
  et~al.}{1994}]{Meszaros_1994}
Meszaros P.,  Rees M.~J.,   Papathanassiou H.,  1994, \mn@doi [The
  Astrophysical Journal] {10.1086/174559}, 432, 181

\bibitem[\protect\citeauthoryear{Minaev \& Pozanenko}{Minaev \&
  Pozanenko}{2019}]{Minaev_2019}
Minaev P.~Y.,  Pozanenko A.~S.,  2019, \mn@doi [Monthly Notices of the Royal
  Astronomical Society] {10.1093/mnras/stz3611}, 492, 1919–1936

\bibitem[\protect\citeauthoryear{Murase}{Murase}{2008}]{Murase:2008sp}
Murase K.,  2008, \mn@doi [Phys. Rev. D] {10.1103/PhysRevD.78.101302}, 78,
  101302

\bibitem[\protect\citeauthoryear{Murase, Ioka, Nagataki  \& Nakamura}{Murase
  et~al.}{2008}]{Murase:2008mr}
Murase K.,  Ioka K.,  Nagataki S.,   Nakamura T.,  2008, \mn@doi [Phys. Rev. D]
  {10.1103/PhysRevD.78.023005}, 78, 023005

\bibitem[\protect\citeauthoryear{Mészáros}{Mészáros}{2015}]{https://doi.org/10.48550/arxiv.1511.01396}
Mészáros P.,  2015, Gamma Ray Bursts as Neutrino Sources,
  \mn@doi{10.48550/ARXIV.1511.01396}, \url {https://arxiv.org/abs/1511.01396}

\bibitem[\protect\citeauthoryear{Mücke, Engel, Rachen, Protheroe  \&
  Stanev}{Mücke et~al.}{2000}]{2000muck}
Mücke A.,  Engel R.,  Rachen J.,  Protheroe R.,   Stanev T.,  2000, \mn@doi
  [Computer Physics Communications] {10.1016/s0010-4655(99)00446-4}, 124,
  290–314

\bibitem[\protect\citeauthoryear{Nava, Ghirlanda, Ghisellini  \& Firmani}{Nava
  et~al.}{2008}]{Nava2008}
Nava L.,  Ghirlanda G.,  Ghisellini G.,   Firmani C.,  2008, \mn@doi [Monthly
  Notices of the Royal Astronomical Society]
  {10.1111/j.1365-2966.2008.13758.x}, 391, 639

\bibitem[\protect\citeauthoryear{{Oganesyan}, {Nava}, {Ghirlanda}  \&
  {Celotti}}{{Oganesyan} et~al.}{2017}]{Oganesyan17}
{Oganesyan} G.,  {Nava} L.,  {Ghirlanda} G.,   {Celotti} A.,  2017, \mn@doi
  [\apj] {10.3847/1538-4357/aa831e}, \href
  {https://ui.adsabs.harvard.edu/abs/2017ApJ...846..137O} {846, 137}

\bibitem[\protect\citeauthoryear{{Oganesyan}, {Nava}, {Ghirlanda}, {Melandri}
  \& {Celotti}}{{Oganesyan} et~al.}{2019}]{Oganesyan19}
{Oganesyan} G.,  {Nava} L.,  {Ghirlanda} G.,  {Melandri} A.,   {Celotti} A.,
  2019, \mn@doi [\aap] {10.1051/0004-6361/201935766}, \href
  {https://ui.adsabs.harvard.edu/abs/2019A&A...628A..59O} {628, A59}

\bibitem[\protect\citeauthoryear{{Petropoulou} \& {Mastichiadis}}{{Petropoulou}
  \& {Mastichiadis}}{2012}]{PM12}
{Petropoulou} M.,  {Mastichiadis} A.,  2012, \mn@doi [\mnras]
  {10.1111/j.1365-2966.2012.20460.x}, \href
  {https://ui.adsabs.harvard.edu/abs/2012MNRAS.421.2325P} {421, 2325}

\bibitem[\protect\citeauthoryear{Petropoulou \& Mastichiadis}{Petropoulou \&
  Mastichiadis}{2018}]{PM18}
Petropoulou M.,  Mastichiadis A.,  2018, \mn@doi [Monthly Notices of the Royal
  Astronomical Society] {10.1093/mnras/sty833}, 477, 2917

\bibitem[\protect\citeauthoryear{{Petropoulou}, {Dimitrakoudis}, {Mastichiadis}
   \& {Giannios}}{{Petropoulou} et~al.}{2014}]{14MPGM}
{Petropoulou} M.,  {Dimitrakoudis} S.,  {Mastichiadis} A.,   {Giannios} D.,
  2014, \mn@doi [\mnras] {10.1093/mnras/stu1362}, \href
  {https://ui.adsabs.harvard.edu/abs/2014MNRAS.444.2186P} {444, 2186}

\bibitem[\protect\citeauthoryear{Pitik, Tamborra  \& Petropoulou}{Pitik
  et~al.}{2021}]{pitik21}
Pitik T.,  Tamborra I.,   Petropoulou M.,  2021, \mn@doi [Journal of Cosmology
  and Astroparticle Physics] {10.1088/1475-7516/2021/05/034}, 2021, 034

\bibitem[\protect\citeauthoryear{{Preece}, {Briggs}, {Mallozzi}, {Pendleton},
  {Paciesas}  \& {Band}}{{Preece} et~al.}{1998}]{Preece98}
{Preece} R.~D.,  {Briggs} M.~S.,  {Mallozzi} R.~S.,  {Pendleton} G.~N.,
  {Paciesas} W.~S.,   {Band} D.~L.,  1998, \mn@doi [\apjl] {10.1086/311644},
  \href {https://ui.adsabs.harvard.edu/abs/1998ApJ...506L..23P} {506, L23}

\bibitem[\protect\citeauthoryear{{Racusin} et~al.,}{{Racusin}
  et~al.}{2008}]{2008Racusin}
{Racusin} J.~L.,  et~al., 2008, \mn@doi [\nat] {10.1038/nature07270}, \href
  {https://ui.adsabs.harvard.edu/abs/2008Natur.455..183R} {455, 183}

\bibitem[\protect\citeauthoryear{Razzaque}{Razzaque}{2010}]{Razzaque_2010}
Razzaque S.,  2010, \mn@doi [The Open Astronomy Journal]
  {10.2174/1874381101003010150}, 3, 150–155

\bibitem[\protect\citeauthoryear{Rees \& Meszaros}{Rees \&
  Meszaros}{1994}]{Rees_1994}
Rees M.~J.,  Meszaros P.,  1994, \mn@doi [The Astrophysical Journal]
  {10.1086/187446}, 430, L93

\bibitem[\protect\citeauthoryear{{Sari}, {Narayan}  \& {Piran}}{{Sari}
  et~al.}{1996}]{1996Sari-Nar-Piran}
{Sari} R.,  {Narayan} R.,   {Piran} T.,  1996, \mn@doi [\apj] {10.1086/178136},
  \href {https://ui.adsabs.harvard.edu/abs/1996ApJ...473..204S} {473, 204}

\bibitem[\protect\citeauthoryear{Sari, Piran  \& Narayan}{Sari
  et~al.}{1998}]{Sari_1998}
Sari R.,  Piran T.,   Narayan R.,  1998, \mn@doi [The Astrophysical Journal]
  {10.1086/311269}, 497, L17

\bibitem[\protect\citeauthoryear{Schnabel, Gal  \& Aly}{Schnabel
  et~al.}{2021}]{schnabel2021km3net}
Schnabel J.,  Gal T.,   Aly Z.,  2021, The KM3NeT Open Science System
  (\mn@eprint {arXiv} {2101.06751})

\bibitem[\protect\citeauthoryear{Toma, Wu  \& Mészáros}{Toma
  et~al.}{2011}]{PHIS}
Toma K.,  Wu X.-F.,   Mészáros P.,  2011, \mn@doi [Monthly Notices of the
  Royal Astronomical Society] {10.1111/j.1365-2966.2011.18807.x}, 415, 1663

\bibitem[\protect\citeauthoryear{Totani}{Totani}{1998}]{Totani_1998}
Totani T.,  1998, \mn@doi [The Astrophysical Journal] {10.1086/311772}, 509,
  L81

\bibitem[\protect\citeauthoryear{Tu \& Wang}{Tu \& Wang}{2018}]{Tu_2018}
Tu Z.~L.,  Wang F.~Y.,  2018, \mn@doi [The Astrophysical Journal]
  {10.3847/2041-8213/aaf4b8}, 869, L23

\bibitem[\protect\citeauthoryear{Vietri}{Vietri}{1995}]{Vietri_1995}
Vietri M.,  1995, \mn@doi [The Astrophysical Journal] {10.1086/176448}, 453,
  883

\bibitem[\protect\citeauthoryear{Vietri}{Vietri}{1997}]{Vietri}
Vietri M.,  1997, \mn@doi [Phys. Rev. Lett.] {10.1103/PhysRevLett.78.4328}, 78,
  4328

\bibitem[\protect\citeauthoryear{Waxman}{Waxman}{1995}]{Waxman_1995}
Waxman E.,  1995, \mn@doi [The Astrophysical Journal] {10.1086/309715}, 452

\bibitem[\protect\citeauthoryear{Waxman \& Bahcall}{Waxman \&
  Bahcall}{1997}]{waxman97b}
Waxman E.,  Bahcall J.,  1997, \mn@doi [Phys. Rev. Lett.]
  {10.1103/PhysRevLett.78.2292}, 78, 2292

\makeatother
\end{thebibliography}




\bsp	
\label{lastpage}
\end{document}